\documentclass[%
reprint,
superscriptaddress,
amsmath,
amssymb,
aps,
twocolumn,
longbibliography,
pra,
floatfix
]{revtex4-1}

\usepackage{xcolor}
\usepackage[colorlinks,linkcolor=blue,citecolor=blue,urlcolor=blue]{hyperref} 

\usepackage{graphicx}
\usepackage[english]{babel}
\usepackage[utf8]{inputenc}
\usepackage{amssymb}
\usepackage{tikz}
\usepackage{pgfplots}
\usepgfplotslibrary{colormaps}
\usetikzlibrary{pgfplots.colormaps} 
\usetikzlibrary[pgfplots.colormaps] 
\usepackage[capitalize]{cleveref}
\usepackage{braket}
\usepackage{siunitx}
\usepackage{color}
\usepackage{csquotes}
\usepackage{newtxtext}
\usepackage{newtxmath}
\usepackage{soul}
\usepackage{shellesc}

\usetikzlibrary{external}
\usetikzlibrary{shapes.geometric}
\usetikzlibrary{arrows.meta,arrows}

\newcommand{\secref}[1] {Sec.~\ref{#1}}
\newcommand{\figsref}[1] {Figs.~\ref{#1}}

\makeatletter 
\renewcommand{\fnum@figure}{FIG~\thefigure}
\makeatother

\begin{document}
\title{Regimes of atomic diffraction: Raman versus Bragg diffraction in retroreflective geometries\\[1ex]
	\normalsize\normalfont{Published in \href{https://journals.aps.org/pra/abstract/10.1103/PhysRevA.101.053610}{Phys. Rev. A {\bfseries 101}, 053610 (2020)}}}
\author{Sabrina Hartmann}
\email{sabrina.hartmann@uni-ulm.de}
\affiliation{%
	Institut für Quantenphysik and Center for Integrated Quantum Science and Technology (IQ\textsuperscript{ST}), Universität Ulm, Albert-Einstein-Allee 11, D-89081 Ulm, Germany
}%
\author{Jens Jenewein}%
\email{jens.jenewein@uni-ulm.de}
\affiliation{%
	Institut für Quantenphysik and Center for Integrated Quantum Science and Technology (IQ\textsuperscript{ST}), Universität Ulm, Albert-Einstein-Allee 11, D-89081 Ulm, Germany
}%

\author{Enno Giese}%
\affiliation{%
	Institut für Quantenphysik and Center for Integrated Quantum Science and Technology (IQ\textsuperscript{ST}), Universität Ulm, Albert-Einstein-Allee 11, D-89081 Ulm, Germany
}%

\author{Sven Abend}%
\affiliation{%
	Institut für Quantenoptik, Leibniz Universität Hannover, Welfengarten 1, D-30167 Hannover, Germany
}%

\author{Albert Roura}%
\affiliation{%
	Institute of Quantum Technologies, German Aerospace Center (DLR), Söflinger Str.~100, D-89077 Ulm, Germany 
}

\author{Ernst~M. Rasel}%
\affiliation{%
	Institut für Quantenoptik, Leibniz Universität Hannover, Welfengarten 1, D-30167 Hannover, Germany
}%

\author{Wolfgang~P. Schleich}%
\affiliation{%
	Institut für Quantenphysik and Center for Integrated Quantum Science and Technology (IQ\textsuperscript{ST}), Universität Ulm, Albert-Einstein-Allee 11, D-89081 Ulm, Germany
}%
\affiliation{%
	Institute of Quantum Technologies, German Aerospace Center (DLR), Söflinger Str.~100, D-89077 Ulm, Germany 
}%
\affiliation{%
	Hagler Institute for Advanced Study and Department of Physics and Astronomy, Institute for Quantum Science and Engineering (IQSE), Texas A\&M AgriLife Research, Texas A\&M University, College Station, TX 77843-4242 USA
}%

\date{\today}

\begin{abstract}
We provide a comprehensive study of atomic Raman and Bragg diffraction when coupling to a pair of counterpropagating light gratings (double diffraction) or to a single one (single diffraction) and discuss the transition from one case to the other in a retroreflective geometry as the Doppler detuning changes.
In contrast to single diffraction, double Raman loses its advantage of high diffraction efficiency for short pulses and has to be performed in a Bragg-type regime.
Moreover, the structure of double diffraction leads to further limitations for broad momentum distributions on the efficiency of mirror pulses, making the use of (ultra) cold ensembles essential for high diffraction efficiency. 
\end{abstract}
\maketitle

\section{\label{sec:Introduction}Introduction}
\noindent
Atoms diffracted by optical gratings in a retroreflective setup display double diffraction in two directions if there is no initial Doppler detuning~\cite{enhancing_area_leveque_2009,malossi_double_2010,ahlers_double_2016,kuber2016experimental,Giese,pagel_bloch_2019,gebbe_twin-lattice_2019}.

Together with other large-momentum transfer techniques~\cite{muller_atom_2008,chiow_$102ensuremathhbark$_2011,PhysRevA.88.053620,kovachy2015quantum,abend_atom-chip_2016,gebbe_twin-lattice_2019}, this process is not only important for earth-bound light-pulse atom interferometers~\cite{kasevich_atomic_1991,RevModPhys.81.1051,kleinert_representation-free_2015,bongs2019taking} in the horizontal direction, as proposed for gravitational wave detectors with horizontal baselines~\cite{Hogan2011,
	Schubert2019,canuel2019elgar}, but also for interferometers under microgravity conditions~\cite{muntinga_interferometry_2013,geiger2011detecting,becker_space-borne_2018,elliott_nasas_2018}.
Double-diffraction schemes can be implemented with both Raman~\cite{kasevich_atomic_1991,enhancing_area_leveque_2009} and Bragg~\cite{torii_mach-zehnder_2000,ahlers_double_2016} processes that require different components.
The design of ambitious atom-interferometric experiments, such as future space missions~\cite{Aguilera_2014,frye_bose-einstein_2019, tino2019sage, berg2019exploring,el-neaj_aedge_2019-1,frye_bose-einstein_2019},  must therefore include specifications for the diffraction mechanism.
We present in this article a detailed study of Raman and Bragg diffraction with particular emphasis on microgravity conditions or horizontal configurations based on retroreflective setups. 

In light-pulse atom interferometry beam splitters and mirrors are realized by the diffraction of (ultra) cold atoms from light waves. Such interferometers constitute precise inertial sensors that measure the atomic motion with respect to a reference. In many setups the light wave is retroreflected~\cite{peters2001high} by a mirror that constitutes such a reference. However, because a pair of different frequencies is necessary in general to diffract atoms of arbitrary velocities, such a geometry naturally leads to two counterpropagating diffraction gratings. Accelerations prior to the light-pulse, such as those due to gravity, cause a Doppler detuning with respect to one of the gratings, and result in effectively single diffraction. However, in a horizontally aligned setup or under microgravity conditions the result of such a smaller initial acceleration is often more complex. For an insufficient Doppler detuning the diffraction process changes drastically so that both gratings are relevant and diffract in opposite directions. In this article we therefore study the difference between Raman and Bragg as well as their behavior in single and double diffraction~\cite{enhancing_area_leveque_2009,malossi_double_2010,ahlers_double_2016,kuber2016experimental,Giese,pagel_bloch_2019,gebbe_twin-lattice_2019}.

To compare the different diffraction techniques we focus on the efficiency, which is determined by the duration of the pulse and the width of the atomic momentum distribution. The pulse duration determines the regime of diffraction, which is in turn connected to a particular velocity selectivity. In principle Raman and Bragg diffraction can be operated in different regimes, so that even thermal atoms~\cite{PhysRevLett.114.063002,PhysRevA.90.033608} can be diffracted by Raman whereas sub-recoil-cooled atoms are essential for Bragg. We compare the regimes of Raman and Bragg as well as single and double diffraction, study the resulting resonance width and show which Doppler detunings are necessary for single diffraction. We find that the border between Raman and Bragg blurs depending on the parameter regime.

We compare and contrast Raman and Bragg diffraction in \secref{sec:DiffractionMechanisms} and explain in \secref{sec:NumericalTreatment} how to numerically solve the corresponding differential equations. In \secref{sec:lResonanceWidth}, we discuss the width of the resonance in momentum space. Furthermore, we calculate respectively the diffraction efficiency and losses in \secref{sec:DiffractionEfficiency} and \secref{sec:DiffractionLosses}. These losses can be substantially high unless sufficiently narrow momentum distributions are employed. In \secref{sec:Interferometer} we show the effect of losses on the amplitude of a Mach-Zehnder interferometer signal and its contrast. Finally, in \secref{sec:Transition} we discuss the role of the Doppler detuning in the transition from double to single diffraction in a retroreflective setup. We conclude in \secref{sec:Conclusion} by briefly summarizing our results and providing an outlook. Additionally, we provide in \cref{App:DGLs} the general differential equations which allow to describe a multitude of diffraction processes for arbitrary resonance conditions. \cref{app:Wavefunctions} gives the explicit expressions to calculate the main path of a wave packet through a Mach-Zehnder interferometer.

\section{\label{sec:DiffractionMechanisms} Raman vs. Bragg diffraction}
\begin{figure}[t]
	\includegraphics[width = \columnwidth]{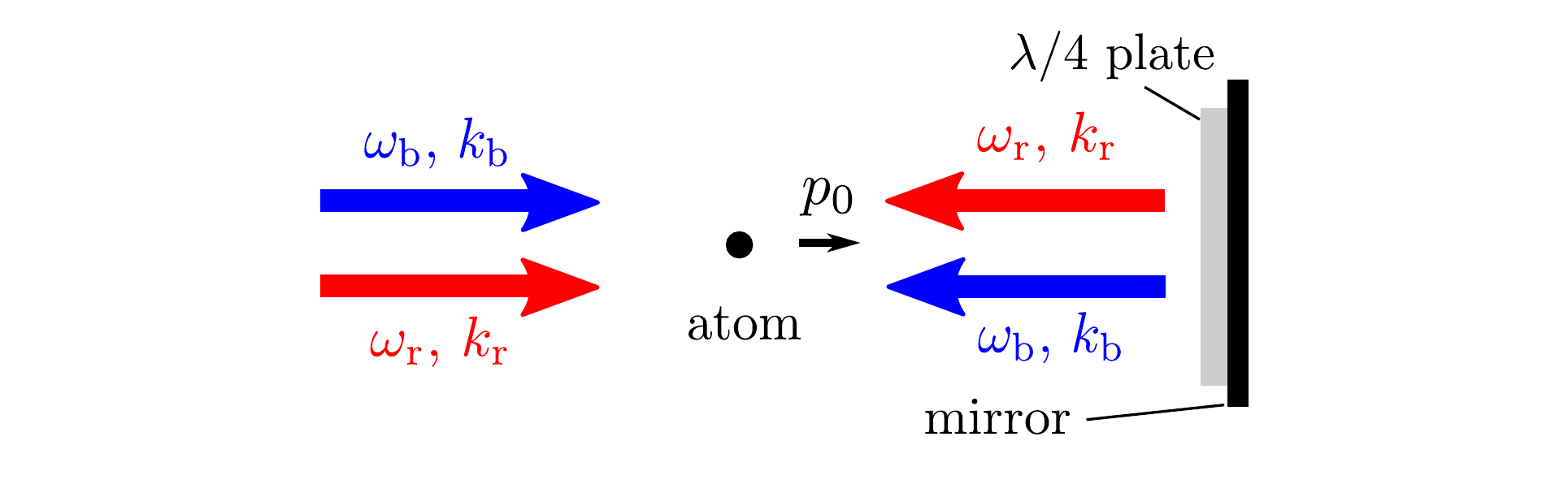}
	\caption{Schematic setup of single diffraction in a retroreflective geometry, realized by a mirror and a $\lambda / 4$ wave plate. The atom with initial momentum $p_0$ interacts with only one of the two counterpropagating optical gratings since the other one is Doppler-detuned. Each grating consists of two  light fields with frequencies  $\omega_\mathrm{b}$ and $\omega_\mathrm{r}$, whose corresponding wave numbers are $k_\mathrm{b}$ and $k_\mathrm{r}$, respectively.}
	\label{fig:Retroreflection}
\end{figure}
An atom interacting with periodic light structures far detuned from the atomic resonance is diffracted by photon absorption and subsequent stimulated emission. In the following, we discuss two \textit{diffraction mechanisms}: Raman and Bragg. Their main difference is that in Raman diffraction the internal state is changed during the two-photon process, while in Bragg diffraction it remains unaffected. Consequently, the two frequencies $\omega_\mathrm{b}$ and $\omega_\mathrm{r}$ that generate the diffraction gratings have to be adjusted to drive the process and thus, their difference $\Delta \omega = \omega_\mathrm{b} - \omega_\mathrm{r}$ is mechanism dependent.

Conventionally, these two different counterpropagating light fields with detuned frequencies are generated using a retroreflective setup, see \cref{fig:Retroreflection}. They are usually derived from the same source, guided through common optics to the setup and retroreflected at the other side of the atomic sample. This way, two optical lattices are formed that propagate in opposite directions. To avoid spurious standing waves, orthogonal polarizations are chosen and turned upon retroreflection through a $\lambda/4$ plate ~\cite{enhancing_area_leveque_2009,ahlers_double_2016}. In general, one can distinguish between two different \emph{diffraction geometries}: (i) \emph{Single diffraction}, where only one of the two gratings dominates the dynamics, and (ii) \emph{double diffraction}, where both gratings are relevant for the diffraction process. In the following, we discuss the difference between both cases.

\subsection{Single-diffraction geometry}
\begin{figure}[htb]
	\includegraphics[width =\columnwidth]{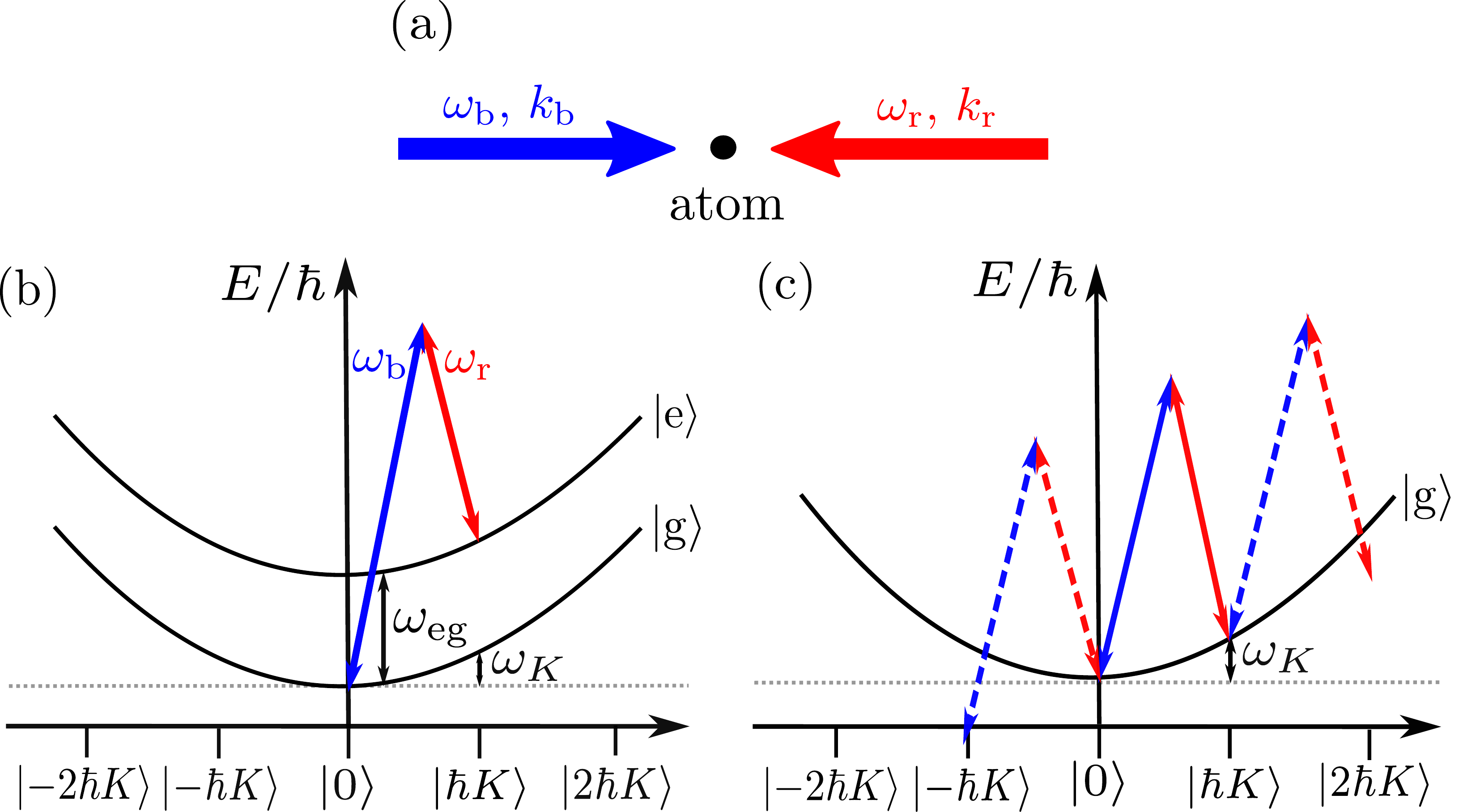}
	\caption{Single Raman versus single Bragg diffraction. The schematic setup  (a) shows an atom interacting with two counterpropagating light fields of frequencies $\omega_\mathrm{b}$ and $\omega_\mathrm{r}$. A photon from a plane wave with frequency $\omega_\mathrm{b}$ is absorbed, while a subsequent photon with frequency $\omega_\mathrm{r}$ is emitted in the opposite direction causing a total recoil of $\hbar K \equiv \hbar (k_{\mathrm b} + k_{\mathrm r})$. During the process, the atom gains kinetic energy $\hbar \omega_K$, the recoil energy. The energy-momentum diagram for single Raman diffraction (b) shows that this process is resonant if the energy difference $\hbar(\omega_\mathrm{b} - \omega_\mathrm{r})$ between the light fields is equal to the recoil energy in addition to the energy difference $\hbar \omega_\mathrm{eg}$ between the internal ground state $\Ket{\mathrm g}$ and excited state $\Ket{ \mathrm e}$. For single Bragg diffraction (c), the process is resonant if $\hbar(\omega_\mathrm{b} - \omega_\mathrm{r})$ is equal to the recoil energy. The dashed lines denote off-resonant higher-order transitions.}
	\label{fig:SingleDiffraction}
\end{figure}

For single diffraction in a retroreflective setup, the atom has an initial momentum that is larger or in the order of the recoil experienced during the diffraction process. In this case, only one of the two counterpropagating gratings is resonant, whereas the other one is strongly Doppler-detuned and becomes irrelevant for the diffraction process. For an atom initially at rest, no Doppler detuning arises and double diffraction occurs. We show numerically in \secref{sec:Transition} the transition from double to single diffraction by increasing the initial momentum and by that the Doppler detuning of one gratings, while keeping the other resonant. If the setup is vertical and therefore parallel to gravity in earth-based experiments, the atoms are naturally accelerated after the release from a trap or launched upwards e.g. in an atomic fountain, so that such a situation conventionally leads to a Doppler-detuned grating.

Since in single diffraction only one of the gratings dominates the process, a simplified model that highlights the relevant physical principles is conventionally used to describe the dynamics~\cite{muller_atom-wave_2008,szigeti_why_2012}. It consists only of one single optical grating, i.e. two counterpropagating laser beams, see~\cref{fig:SingleDiffraction}(a). If this corresponds to the experimental situation, single diffraction can be also performed for an atom initially at rest with an initial momentum of $p_0 = 0$. In the following, we focus on this case, but emphasize that a retroreflective setup with strong Doppler detuning leads to the same results.

The two counterpropagating beams that form the relevant diffraction grating have frequencies $\omega_{\mathrm{b}}$ and $\omega_{\mathrm{r}}$, see~\cref{fig:SingleDiffraction}(a). The atom absorbs a photon with momentum $\hbar k_\mathrm{b}$ and emits a photon with momentum $-\hbar k_\mathrm{r}$ in the opposite direction. Here, $k_\mathrm{b}$ and $k_\mathrm{r}$ are the wave numbers of the corresponding light fields. This two-photon process leads to a total momentum transfer of $\hbar K \equiv \hbar k_\mathrm{b} + \hbar k_\mathrm{r}$ as a consequence of momentum conservation. At the same time, the energy $\hbar \Delta \omega$ is absorbed. Energy conservation gives rise to different resonance conditions, depending on the diffraction mechanism. In the following, we discuss the two mechanisms and their respective resonance conditions.

\subsubsection{Single Raman diffraction}
In single Raman diffraction the internal state changes during the two-photon process from the ground state $\ket{ \mathrm g}$ to the excited state $\ket{\mathrm e}$ or vice versa, see \cref{fig:SingleDiffraction}(b). Thus, the energy difference $\hbar \omega_{\mathrm{eg}}$ between these states influences the resonance condition. A transition is resonant if it begins and ends on one of the parabolae representing the kinetic energy of each internal state. For a resonant process, energy \textit{and} momentum have to be conserved, which requires the energy difference $\hbar \Delta \omega$ to be equal to the kinetic energy $\hbar \omega_K$ gained through the momentum transfer plus the energy difference of the internal states $\hbar \omega_{\mathrm{eg}}$. Hence, the resonance condition can be written as
\begin{equation}
\Delta \omega = \omega_K + \omega_{\mathrm{eg}} + \omega_{\mathrm{AC}}
\label{eq:ResCondRaman}
\end{equation}
with the AC Stark shift $\omega_{\mathrm{AC}}$
and the recoil frequency 
\begin{equation}
\omega_K \equiv \frac{\hbar K^2}{2M},
\end{equation}
where $M$ is the mass of the atom. 

Since the AC Stark shift can be compensated, we choose in the following $\omega_{\mathrm{AC}} = 0$. In principle, the mean initial momentum of the atom enters the resonance condition as well but we have chosen the initial momentum $p_0 = 0$ for an atom initially at rest. The choice of $\Delta \omega$ compatible with $p_0 \neq 0$ will be discussed in \secref{sec:Transition}.

Within the rotating wave approximation, only terms that oscillate with a frequency significantly lower than the involved optical frequencies have to be taken into account. By appropriately choosing the laser frequencies, the large internal frequency difference $\omega_\text{eg}$ between the two states is canceled following the resonance condition from~\cref{eq:ResCondRaman}. In addition, higher-order transitions are strongly suppressed since they are detuned by $\hbar \omega_{\mathrm{eg}}$ and can be adiabatically eliminated~\cite{bernhardt_coherent_1981,marte_multiphoton_1992,brion_adiabatic_2007} as further discussed in \cref{App:DGLs}. Consequently, $\ket{ \mathrm g}$ and $\ket{\mathrm e}$ couple like an effective two-level system which, according to Ref.~\cite{moler}, is given by
\begin{equation}
\begin{pmatrix} \dot g_n \\ \dot e_{n+1} \end{pmatrix} \! =\mathrm i \Omega(t) \! \begin{pmatrix} 0 &  \mathrm e^{-\mathrm i (\omega_\mathrm{D} +2  n\omega_K) t}  \\  \mathrm e^{\mathrm i( \omega_\mathrm{D} +2 \mathrm n\omega_K )t} & 0  \end{pmatrix}   \! \begin{pmatrix}  g_n \\  e_{n+1} \end{pmatrix}
\label{eq:SingleRaman}
\end{equation}
where we already have used the resonance condition from~\cref{eq:ResCondRaman} and consider vanishing laser phases. A more general form for all geometries and mechanisms can be found in~\cref{App:DGLs}. The differential equations are formulated in an interaction picture with respect to the free evolution. 

\Cref{eq:SingleRaman} describes Rabi oscillations~\cite{bernhardt_coherent_1981} between the probability amplitudes $g_n \equiv g(p + n \hbar K)$ and $e_n \equiv e(p + n \hbar K)$ of the ground and excited state in momentum representation. The coupling strength $\Omega = \Omega(t)$ depends on the intensity and the pulse shape of the grating. The considered momentum state is denoted by the index $n$. The process shown in~\cref{fig:SingleDiffraction} is resonant for $n = 0$ so that the exponent $n \omega_K$ vanishes. The Doppler frequency 
\begin{equation}
\omega_\mathrm{D}(p) = \frac{p K}{M}
\end{equation}
acts as a detuning for a momentum distribution around a resonant momentum $p_0=0$. The distribution is diffracted more efficiently the smaller $\omega_\mathrm{D}$ is, leading to velocity selectivity~\cite{PhysRevLett.66.2297,kozuma_coherent_1999,debs_cold-atom_2011}. The system of differential equations in~\cref{eq:SingleRaman} is closed and can be solved analytically for box-shaped pulses.

\subsubsection{Single Bragg diffraction}
\label{subsec:SingleBraggDiffraction}

In single Bragg diffraction the internal state is not changed during the process as depicted in~\cref{fig:SingleDiffraction}(c), which requires the modification of the resonance condition to
\begin{equation}
\Delta \omega = \omega_K
\label{eq:ResCondBragg}
\end{equation}
so that the transferred energy solely corresponds to the gained kinetic energy. 

The system of differential equations according to Ref.~\cite{Giese} takes the form 
\begin{equation}
\begin{split}
\mathrm i \dot g_n = &- \Omega(t) \, \mathrm e^{- \mathrm i \omega_\mathrm{D} t} \, g_{n + 1} \, \mathrm e^{- 2 \mathrm i n \omega_K  t} \\
&- \Omega(t) \, \mathrm e^{\mathrm i \omega_\mathrm{D} t} \, g_{n - 1} \, \mathrm e^{2 \mathrm i (n - 1) \omega_K t}.
\end{split}
\label{eq:SingleBragg}
\end{equation}
In contrast to single Raman, higher-order diffraction is possible since these off-resonant transitions are detuned in the order of $\omega_K$, and not by $\omega_{\mathrm{eg}}$, which is more than five orders of magnitude larger. Such transitions are denoted by dashed lines in~\cref{fig:SingleDiffraction}(c). These off-resonant higher orders are described in~\cref{eq:SingleBragg} by terms which oscillate with $n \omega_K$ and disturb the diffraction processes. They are prominent in the Raman-Nath (Kapitza-Dirac) regime~\cite{gould_diffraction_1986,muller_atom-wave_2008} where the pulse durations are short and intensities are high, so that $\epsilon \equiv \Omega/\omega_K \gtrsim 1 $. These effects can be understood as a manifestation of energy-time uncertainty. For $\epsilon \lesssim 1$ their effect decreases and they are suppressed in the \textit{Bragg regime}, where $\epsilon \ll 1$. Within this regime, the diffraction process can be treated as an effective two-level system, that undergoes Rabi oscillations. However, for other regimes off-resonant transitions are possible. \Cref{eq:SingleBragg} is not closed anymore and analytical solutions cannot be obtained, so we have to treat it numerically even for box-shaped pulses. The Doppler frequency $\omega_\mathrm{D}$ leads again to velocity selectivity~\cite{szigeti_why_2012}.

 \subsection{Double-diffraction geometry}
\begin{figure}[t]
  \includegraphics[width = \columnwidth]{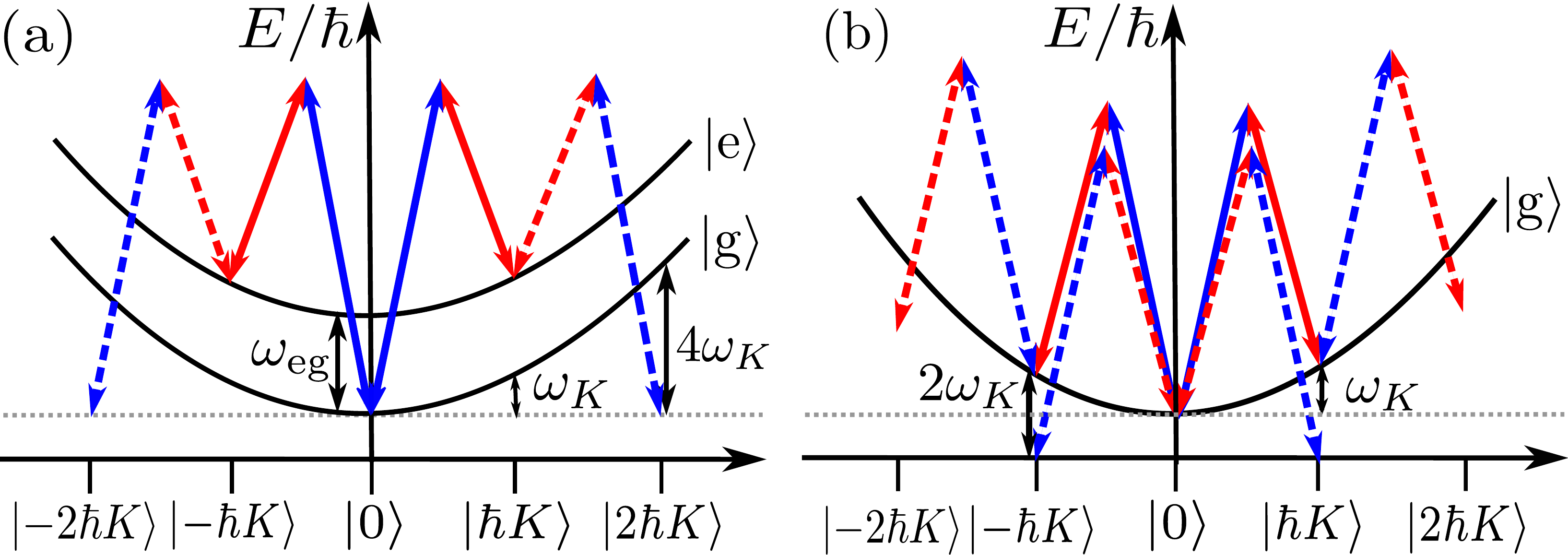}
  \caption{Energy-momentum diagrams for double Raman (a) and double Bragg diffraction (b). The atom interacts with two \textit{pairs} of counterpropagating light fields and gains during this process the momentum $\hbar K$ which can be associated with the recoil frequency $\omega_{K}$. 
  	The two optical gratings allow transitions in opposite directions. Higher-order diffraction is possible for both mechanisms, as shown by dashed lines. They are off-resonant by integer multiples of the recoil energy $\hbar \omega_K$. Here, some of the transitions in Raman diffraction have been omitted, as they are off-resonant by the energy difference $\hbar \omega_\mathrm{eg}$ between ground state $\ket{\mathrm{g}}$ and excited state $\ket{\mathrm{e}}$.}
 \label{fig:DoubleDiffraction}
\end{figure}

In a retroreflective setup and for an atom initially at rest with initial momentum $p_0 = 0$, both counterpropagating optical gratings are equally relevant for the diffraction process, leading to double diffraction. This process is therefore of particular relevance for space missions where the atoms can be naturally released without an initial velocity. But also for earth-based inertial sensing in horizontal direction orthogonal to gravity, the atoms might have no initial momentum and exhibit double diffraction~\cite{ahlers_double_2016,kuber2016experimental}. As already mentioned, different polarizations are used and rotated upon retroreflection by a $\lambda/4$ plate~\cite{enhancing_area_leveque_2009,ahlers_double_2016,kuber2016experimental} to distinguish the two gratings.
The following discussion focuses on perfectly orthogonal polarizations and distinguishable gratings. Polarization imperfections lead to additional couplings not discussed in this article.

\subsubsection{Double Raman diffraction}
\label{subsec:SingleRamanDiffraction}
The atom interacts with two laser pairs and consequently diffraction in both directions is possible as depicted in~\cref{fig:DoubleDiffraction}(a). Compared to single diffraction, the resonance condition does not change. The additional laser pair drives not only a process in the opposite direction, but also off-resonant transitions, which are denoted by the dashed lines in~\cref{fig:DoubleDiffraction}(a). Similar to single Bragg diffraction, their detuning is in the order of $\omega_{K}$ and working in an appropriate regime is required to suppress these transitions. The system of coupled equations is given by
\begin{subequations}
\begin{equation}
\begin{split}
\mathrm i \dot g_n = &- \Omega(t) \, \mathrm e^{-\mathrm i \omega_\mathrm{D} t} \, \mathrm e^{-\mathrm i 2n\omega_K t} \,  e_{n+1} \\
                     &   - \Omega(t) \, \mathrm e^{\mathrm i \omega_\mathrm{D} t} \, \mathrm e^{\mathrm i 2n\omega_K t} \,  e_{n-1}
\end{split}
\end{equation}
\begin{equation}
\begin{split}
\mathrm i \dot e_{n+1} = & -\Omega(t) \, \mathrm e^{-\mathrm i \omega_\mathrm{D} t} \, \mathrm e^{-\mathrm i (4+2n)\omega_K t} \, g_{n+2} \\
                         & -\Omega(t) \, \mathrm e^{\mathrm i \omega_\mathrm{D} t} \, \mathrm e^{\mathrm i 2n\omega_K t} \, g_{n}
\end{split}
\end{equation}
\label{eq:DoubleRaman}
\end{subequations}
\hspace*{-.4em} which is a generalized version of the truncated equations in Ref.~\cite{ThesisLeveque}. In contrast to single Raman, the system is not closed and the equations show that it is possible to diffract into both directions simultaneously. Like in single Bragg diffraction, oscillatory factors in~\cref{eq:DoubleRaman} can be identified by their oscillation with frequency $n \omega_K$. In the appropriate regime, they are suppressed and Rabi oscillations with an effective Rabi frequency of $\sqrt{2} \Omega$ are possible.

\subsubsection{Double Bragg diffraction}
\label{subsec:DoubleBraggDiffraction}
Like in double Raman diffraction, a second laser pair in Bragg also causes transitions in both directions. Since higher orders are not suppressed by a hyperfine splitting, momenta couple resonantly as well as off-resonantly, as shown in \cref{fig:DoubleDiffraction}(b). Additionally, higher-order processes of single Bragg diffraction appear. Therefore, double Bragg diffraction shows the highest complexity with respect to possible transitions. 

Nevertheless, the off-resonant processes can be suppressed when working in the Bragg regime. The system of differential equations ~\cite{Giese}
\begin{equation}
\begin{split}
\mathrm i \dot g_n = & - \Omega(t) \, \mathrm e^{- \mathrm i \omega_\mathrm{D} t} \, g_{n + 1} \, \Big[ \mathrm e^{- 2 \mathrm i (n + 1) \omega_K  t} + \mathrm e^{- 2 \mathrm i n \omega_K t} \Big] \\
                     & - \Omega(t) \, \mathrm e^{\mathrm i \omega_\mathrm{D} t} \, g_{n - 1} \, \Big[ \mathrm e^{2 \mathrm i n \omega_K t} + \mathrm e^{2 \mathrm i (n - 1) \omega_K t} \Big]
\end{split}
\label{eq:DoubleBragg}
\end{equation}
shows that each momentum state is coupled by two transitions that oscillate at different multiples of $\omega_K$. For $n = 0$ and $\omega_\mathrm{D}=0$, we see a simultaneous resonant (time-independent) and off-resonant (oscillating) coupling. Like in double Raman diffraction, Rabi oscillations with an effective Rabi frequency of $\sqrt{2} \Omega$ are possible.

\section{\label{sec:NumericalTreatment}Numerical treatment}
\begin{figure}[t]
	\includegraphics[width=\columnwidth]{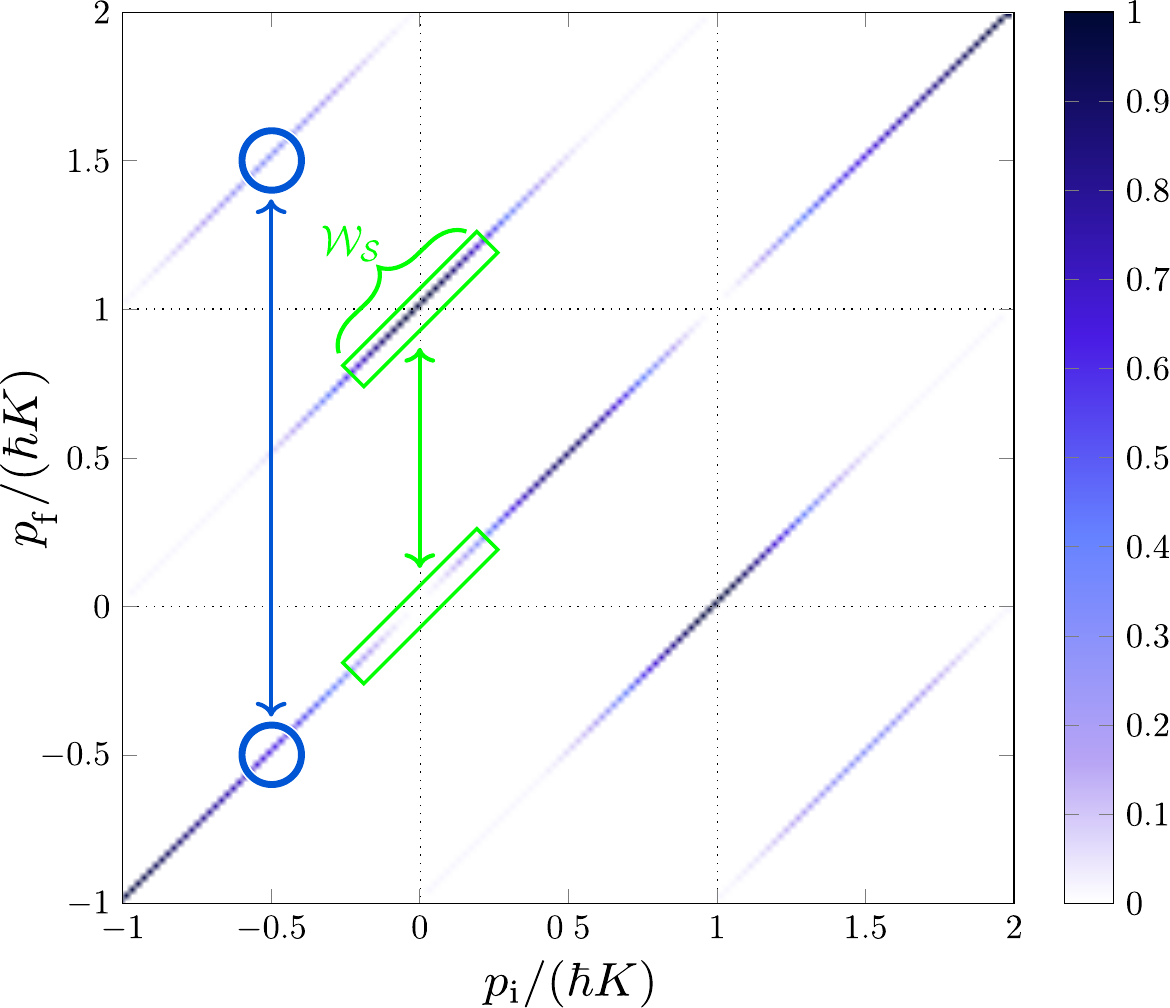}
	\caption{Transition function $\left| G^{(\mathrm{M})}_{\Delta \tau}(p_{\mathrm{f}}, p_{\mathrm{i}})\right|^2$ for a single-Bragg mirror pulse of a Gaussian width $\Delta \tau = \SI{12.5}{\micro \second}$. The diagonals denote the momentum transfer between $p_\mathrm{i}$ and $p_\mathrm{f}$ which is always an integer multiple of $\hbar K$. The process $\Ket{0} \rightarrow \Ket{\hbar K}$ (green rectangles) is resonant. The parameter regime is such that also  second-order processes $\Ket{-0.5 \hbar K} \leftrightarrow \Ket{1.5 \hbar K}$ (blue circles) have a non-vanishing probability. The resonance width $\mathcal{W}_\mathrm{S}$ is defined through the FWHM in \cref{eq:ResonaceWidthSingle}.}
	
	\label{fig:transferMatrix}
\end{figure}
The systems of differential equations for Bragg and Raman diffraction corresponding to \cref{eq:SingleBragg,eq:SingleRaman} as well as \cref{eq:DoubleRaman,eq:DoubleBragg}, are solved numerically with the help of \textsc{Matlab}'s \textsc{ODE45} algorithm. The method is a Runge-Kutta-type algorithm~\cite{shampine_matlab_1997} and we use a relative accuracy of $10^{-3}$ as well as an absolute accuracy of $10^{-6}$. 
Since the systems of differential equation are, with the exception of single Raman, not closed, we have to truncate the range of momenta $\left[-\left(n_\text{max}+1/2\right)\hbar K, \left(n_\text{max}+ 1/2\right)\hbar K\right]$ for the numerical calculation to a finite value of the largest considered diffraction order $n_\text{max}$. If for the computationally most challenging case (i.e. smallest time, greatest detuning or broadest momentum width) the difference between the solution obtained with $n_\text{max}$ and with $n_\text{max}+1$ is at least the same magnitude as the accuracy of our solver algorithm, we assume the effect of the truncation to be negligible. Of course, pulse duration, diffraction mechanism, detuning and considered momentum width influence the truncation.

Throughout the paper, with only one exception [the momentum eigenstates in Fig.~14(b)], we consider as initial state a Gaussian wave packet
$\psi_{\mathrm i}(p_\mathrm{i}) \propto \exp\left[-(p_{\mathrm i} -p_0)^2/(4 \Delta \wp^2 )\right]$ with momentum width $\Delta \wp$ and mean momentum $p_0$.
In this way, we calculate the transition function $G_{\Delta \tau}^{\mathrm{(BS/M)}}(p_{\mathrm{f}},p_{\mathrm{i}})$ connecting the initial wave function $\psi_{\mathrm i} (p_\mathrm{i})$ and the final wave function $\psi_{\mathrm f}(p_\mathrm{f})$ as given in momentum representation by the relation
\begin{equation}
\psi_{\mathrm f}(p_\mathrm{f}) = \int \mathrm d p_\mathrm{i}\,  G_{\Delta \tau}^{\mathrm{(BS/M)}}(p_\mathrm{f},p_\mathrm{i}) \, \psi_{\mathrm i} (p_\mathrm{i}).
\end{equation}
Here we have introduced the superscript BS for a beam splitter and M for a mirror pulse, which in turn depend on the pulse area 
\begin{equation}
	\mathcal A \equiv \int \mathrm d t \, \Omega_\mathrm{R}(t),
\end{equation}
 where we have introduced the effective Rabi frequency
\begin{equation}
\Omega_\mathrm{R}(t) \equiv
\begin{cases}
2 \Omega(t)  & \text{for single diffraction, } \\
\sqrt{2} \Omega(t)  & \text{for double diffraction}  
\end{cases}
\end{equation}
to take the different geometries into account~\cite{giese2015mechanisms}. The value $\mathcal A = \pi / 2$ corresponds to a beam splitter and $\mathcal A = \pi$ to a mirror. We chose the time-dependent coupling strength \mbox{$\Omega(t) \propto \exp\left[-t^2/\left(2 \Delta \tau^2\right)\right]$} to be a Gaussian function of width $\Delta \tau$.

In~\cref{fig:transferMatrix} we show the relevant part of the transition function for a mirror pulse in single Bragg diffraction with pulse duration $\Delta \tau = \SI{12.5}{\micro \second}$, and connecting resonantly the momentum states $\Ket{0}$ and $\Ket{\hbar K}$. Indeed, for a momentum distribution around $p_\mathrm{i}=0$, there exists a high probability to be diffracted to $p_\mathrm{f}=\hbar K$ as indicated by the green boxes. The transition function also shows so-called quasi-resonances~\cite{Giese}, i.e. second-order resonant processes between e.g. $p_\mathrm{i}=-0.5 \hbar K$ and $p_\mathrm{f}=1.5 \hbar K$ displayed by the blue circles.

\begin{figure}[t]
	\includegraphics[width=\linewidth]{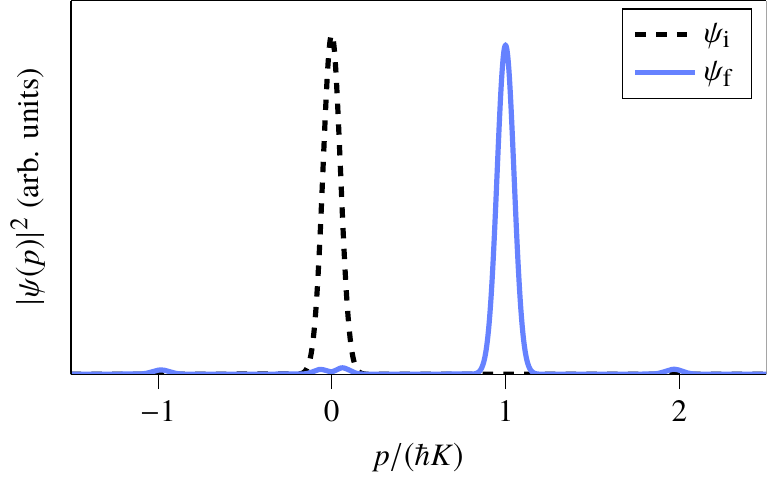}
	\caption{Momentum distribution after a single Bragg diffraction mirror pulse of duration $\Delta \tau = \SI{12.5}{\micro \second}$. The initial distribution $|\psi_{\mathrm i}|^2$ of width $\Delta \wp = 0.05 \hbar K$ is diffracted to the final distribution $|\psi_{\mathrm f}|^2$. In this process not only the resonant momentum state $\Ket{\hbar K}$, but also spurious (i.e. non-resonant) momentum states such as $\Ket{-\hbar K}$ and $\Ket{2 \hbar K}$ are populated. Moreover, some population remains in the initial momentum state $\Ket{0}$.}
	\label{fig:diffractionProcess}
\end{figure}

For short pulse durations such as $\Delta \tau = \SI{12.5}{\micro \second}$, off-resonant higher orders are also populated. We calculate a diffracted momentum distribution with the help of $G_{\Delta \tau}^{\mathrm{(M)}}$ displayed in~\cref{fig:transferMatrix} and show in~\cref{fig:diffractionProcess} how an initial Gaussian momentum distribution of width $\Delta \wp = 0.05 \hbar K$ around $p=0$ (dashed) is diffracted. Most of the diffracted population is centered around $p = \hbar K$, but smaller contributions also appear at other orders, and the initial state is not completely depopulated. 

\section{\label{sec:lResonanceWidth} Width of resonance}
\begin{figure}[htb]
	\includegraphics[width =\linewidth]{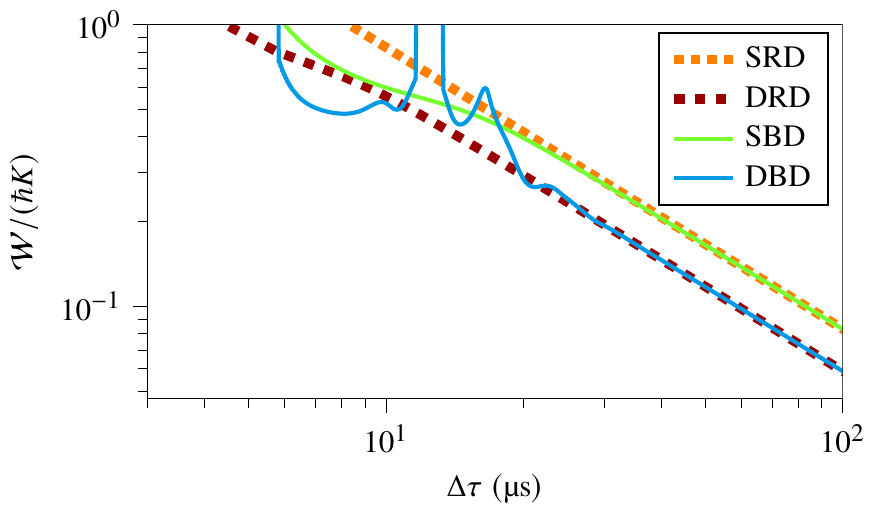}
	\caption{Resonance width $\mathcal W$, defined by \cref{eq:ResonaceWidthSingle,eq:ResonaceWidthDouble}, of single Raman (SRD), double Raman (DRD), single Bragg (SBD) and double Bragg (DBD) diffraction for Gaussian mirror pulses with a varying pulse duration $\Delta \tau$. In general, single diffraction shows always a larger width than double diffraction. For short times, Raman and Bragg differ: Bragg diffraction has a richer structure in this regime, especially double Bragg diffraction. For short pulse durations the initially distinct peaks of the quasi-resonances present in double Bragg start to merge with the central peak causing a large variation of $\mathcal{W}$. For $\mathcal W/(\hbar K)$ larger than unity, the width of the resonance exceeds the separation of the resonant momentum states.}
	\label{fig:ComparisionVelocitySelctivityMirrors}
\end{figure}
\begin{figure*}
	\includegraphics[width =\linewidth]{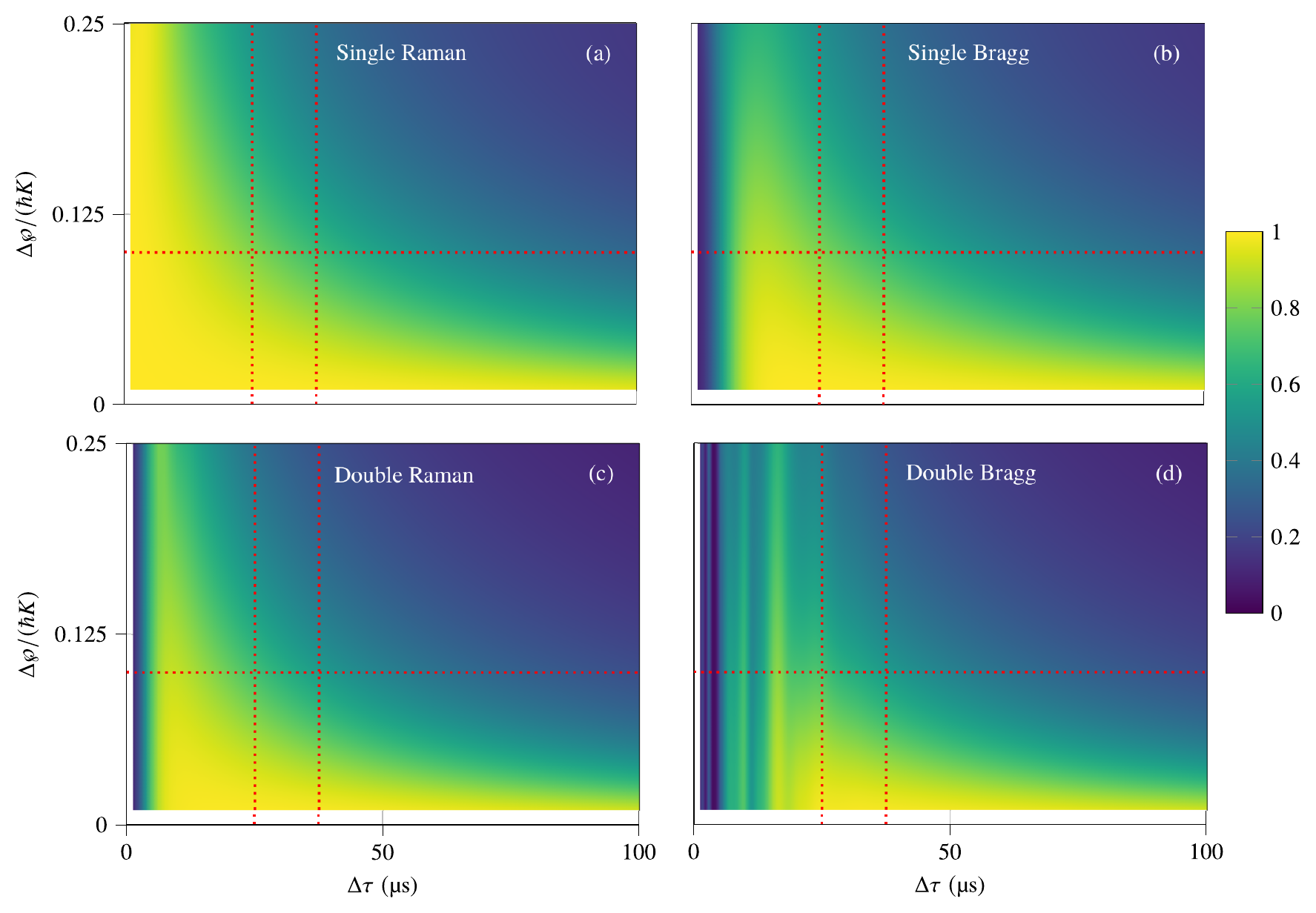}
	\vspace*{-0.1cm}
	\caption{Diffraction efficiency $\mathcal E$ defined by \cref{eq:definitionDiffractionEfficiency} for single (top) and double (bottom) Raman (left) and Bragg (right) mirror pulses. In all four figures we vary the pulse duration $\Delta \tau$ and the momentum width $\Delta \wp$ of the initial state. The efficiency decreases with increasing width and increasing pulse duration. One can see that single Raman diffraction (a) exhibits a high efficiency for the widest parameter range. For single Bragg (b) and double Raman (c) off-resonant higher orders lead to a decrease of efficiency for small $\Delta \tau$. Additional off-resonant couplings lead to a further decrease for double Bragg diffraction (d), while quasi-resonances improve the efficiency for particular durations. The vertical and horizontal dotted lines denote cuts through the three-dimensional distributions and the corresponding curves are presented in \cref{fig:CutDiffractionEfficiency}.}
	\label{fig:ProbabilityGausswithDuration}
\end{figure*}
\begin{figure*}[tb]
	\hspace*{-0.3cm}
	\includegraphics[width =\linewidth]{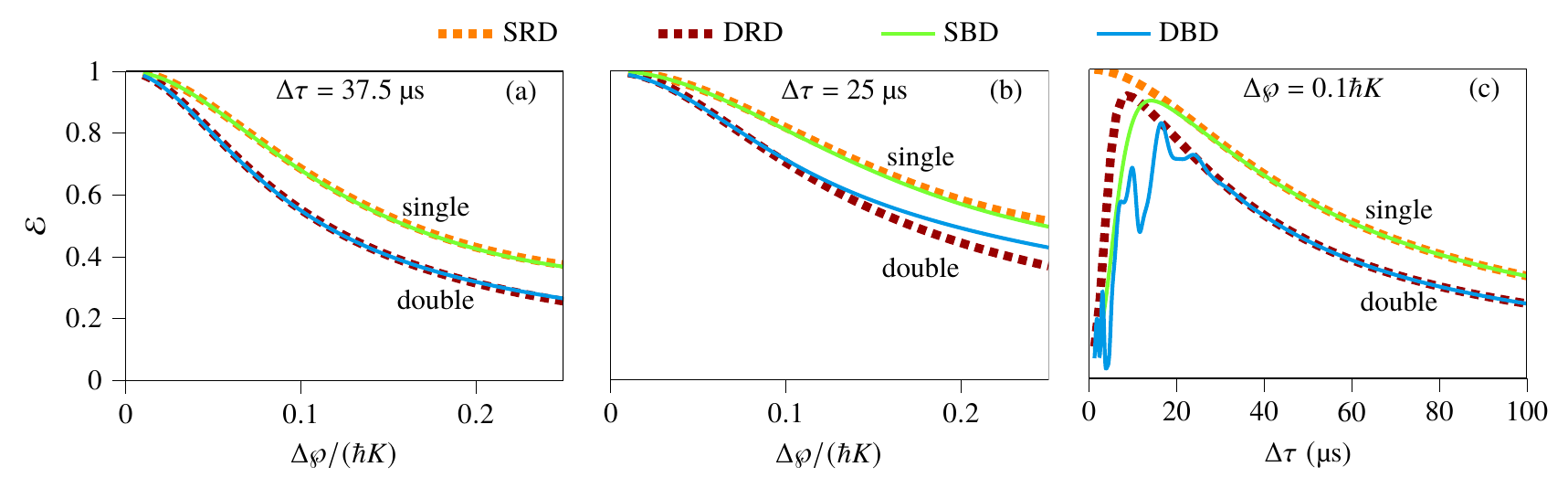}
	\caption{Diffraction efficiency $\mathcal E$ defined by \cref{eq:definitionDiffractionEfficiency} for single and double Raman (SRD and DRD) as well as single and double Bragg (SBD and DBD) mirror pulses.  In (a) we use the pulse duration $\Delta \tau=\SI{37.5}{\micro \second}$, vary the momentum width $\Delta \wp$ and note that the diffraction geometry governs the behavior. In (b) we chose $\Delta \tau =\SI{25}{\micro \second}$, where the efficiency of double Bragg spikes, see \cref{fig:ProbabilityGausswithDuration} (d), due to quasi-resonances. Thus, its efficiency approaches the single-diffraction efficiency for broad distributions. In (c) the pulse duration $\Delta \tau$ is varied, while the momentum width of the initial state is kept at $\Delta \wp = 0.1\hbar K$. Except for single Raman, off-resonant higher orders decrease the efficiency for small $\Delta \tau$.}
	\label{fig:CutDiffractionEfficiency} 
\end{figure*}
The transition function completely determines the diffraction process and contains the resonances, whose width is caused by velocity selectivity. It depends on the parameter regime, i.e. the pulse duration $\Delta \tau$ and the pulse area $\mathcal A$. To characterize these resonances, we calculate the full width at half maximum (FWHM) of the transition function around the resonant momenta marked in~\cref{fig:transferMatrix} by the green brace. The efficiency of the diffraction process is partially determined by the width.

To study the influence of the duration $\Delta \tau$, we calculate the resonance width 
\begin{subequations}
	\begin{equation}
		 \mathcal W_\mathrm{S} \equiv \mathrm{FWHM}_p \bigg[  \Big| G_{\Delta \tau}^{(\mathrm{M})}(p+\hbar K,p) \Big|^2 \bigg]
		 \label{eq:ResonaceWidthSingle} 
	\end{equation}
	for single, and
		\begin{equation}
	  \mathcal W_\mathrm{D} \equiv \mathrm{FWHM}_p \bigg[  \Big| G_{\Delta \tau}^{(\mathrm{M})}(p+\hbar K,p- \hbar K) \Big|^2 \bigg]
	  \label{eq:ResonaceWidthDouble}
	\end{equation}
\end{subequations}
for double diffraction through the FWHM in $p$ around the maximum at $p \approx 0$.

The results for all mechanisms, that is Raman and Bragg, and geometries, that is single and double diffraction, are shown in~\cref{fig:ComparisionVelocitySelctivityMirrors} for mirror pulses with varying  $\Delta \tau$. We note that the resonance for double diffraction is narrower than that for single diffraction. This is a feature of the generalized Rabi oscillations for a three-state system simultaneously driven by the two pairs of counterpropagating laser beams in a double-diffraction setup. As one considers initial momenta slightly away from resonance, the amplitude of the target state for a mirror pulse starts to decrease and this happens faster as a function of the detuning than in single diffraction. This fact is mainly due to an increasing amplitude of diffraction to the intermediate state $\ket{ p + 0\, \hbar K}$ rather than an increasing amplitude of undiffracted atoms in the initial state, as it would be the case for single diffraction. These losses to the intermediate state will be discussed in detail in~\secref{sec:DiffractionLosses}.

In general, both mechanisms (Raman and Bragg) behave the same, but $\mathcal{W}$ depends on the geometry even though all cases display the same scaling. However, for short durations there are differences between the two diffraction mechanisms: especially for Bragg diffraction, off-resonant higher orders and quasi-resonances are far more pronounced in this regime leading to a richer structure. The initially distinct peaks of the quasi-resonances begin to merge with the central peak for decreasing pulse durations and therefore influence the FWHM, so that jumps of $\mathcal{W}$ appear. In this case $\mathcal{W}$ is calculated over all peaks. For small $\Delta \tau$, off-resonant transitions cannot be neglected and couple additionally to the resonant states. Double-Bragg beam-splitter pulses exhibit the same qualitative (but not quantitative) behavior as mirror pulses.

\section{\label{sec:DiffractionEfficiency} Diffraction efficiency}
Next, we analyze the efficiency of the diffraction process defined as
\begin{equation}
\mathcal{E} \equiv \int\limits_{\hbar K/2}^{3 \hbar K /2} \mathrm d p_\mathrm{f} \, \Big| \psi_\mathrm{f}(p_\mathrm{f}) \Big|^2,
\label{eq:definitionDiffractionEfficiency}
\end{equation}
 by integrating over the diffracted momentum distribution\footnote{Alternatively, one could consider the norm for the amplitudes of all momentum eigenstates shifted by $\hbar K$ as a result of the diffraction process. For initial wave packets with momentum widths narrower than $\hbar K$ and centered around $p_0 = 0$, both definitions become equivalent.} around the target momentum $\hbar K$. In contrast to Ref.~\cite{szigeti_why_2012}, which focused on Bragg diffraction, we do not compute the fidelity and thus neglect phases. 
 We perform the simulation for different widths $\Delta \wp$ of the initial Gaussian momentum distribution as well as different pulse durations $\Delta \tau$. Although we present here only the results for mirror pulses, we have found that beam-splitter pulses behave similarly.

Figure~\ref{fig:ProbabilityGausswithDuration} shows the efficiency $\mathcal E$ as a function of $\Delta \wp$ and $\Delta \tau$ for both diffraction mechanisms and geometries. For all cases, an increase of the pulse duration $\Delta \tau$ leads to a decrease of $\mathcal{W}$, and thus to a decrease of the efficiency. Moreover, increasing the width of the initial state $\Delta \wp$ also leads to a decrease of the efficiency since the width of the distribution becomes larger than the width of the resonance.

Figure~\ref{fig:ProbabilityGausswithDuration} compares the efficiency $\mathcal{E}$ of Raman (left) to Bragg (right) and single (top) to double diffraction (bottom). Single Raman in \cref{fig:ProbabilityGausswithDuration}(a) allows the most efficient diffraction for the broadest parameter range. In particular, no efficiency is lost for short pulse durations and Raman diffraction can be performed efficiently in all regimes, even beyond the Bragg-type regime. By comparing the single geometries (a) and (b), we observe for small $\Delta \tau$ a substantial decrease of efficiency in Bragg diffraction. This effect is caused by higher-order diffraction, already discussed in~\secref{subsec:SingleBraggDiffraction} and presented schematically in~\cref{fig:SingleDiffraction}(c). Since off-resonant higher orders appear in double Raman diffraction as well, it is not surprising that we find a similar behavior in \cref{fig:ProbabilityGausswithDuration}(c). In contrast, additionally to higher-order processes, off-resonant couplings are possible in double Bragg diffraction. Figure~\ref{fig:ProbabilityGausswithDuration}(d) confirms this effect since the efficiency decreases faster. However, for small $\Delta \tau$ in double Bragg diffraction we find regions where the efficiency increases due to quasi-resonances. 

In this analysis we have defined the pulse area directly in terms of the coupling strength $\Omega$. However, whenever spurious diffraction orders become relevant, i.e. for parameters beyond the Bragg regime and especially for double Bragg diffraction, the effective Rabi frequency changes \cite{Giese}. Hence, our choice of pulse area might not necessarily lead to the most efficient diffraction. In fact, optimizing the pulse area for each duration and momentum width will increase the efficiency.

For a closer inspection we show in \cref{fig:CutDiffractionEfficiency} three different cuts through the density plot, denoted by the dotted red horizontal and vertical lines in \cref{fig:ProbabilityGausswithDuration}. In~\cref{fig:CutDiffractionEfficiency}(a) we fix the pulse duration to $\Delta \tau = \SI{37.5}{\micro \second}$, where the efficiency for double Bragg does not display a rich structure, and vary the width of the initial state. Raman and Bragg behave similarly and the geometry governs the behavior.

In \cref{fig:CutDiffractionEfficiency}(b) we choose the pulse duration $\Delta \tau = \SI{25}{\micro \second}$. For this duration the efficiency for double Bragg diffraction assumes a maximum as shown in \cref{fig:ProbabilityGausswithDuration}(d). We observe that the diffraction efficiency for double Bragg is larger than for double Raman diffraction if we consider medium and large $\Delta \wp$. This spike is caused by the quasi-resonances discussed previously. They generate side-maxima around the main peak of the transition function and become important for decreasing pulse duration until they merge with the main peak. If the side maxima are large enough, then they lead to an increase of $\mathcal{W}$, which can also be seen in \cref{fig:ComparisionVelocitySelctivityMirrors}. Consequently, broad momentum distributions are diffracted more efficiently, making double Bragg diffraction a possible alternative for thermal atoms.

Finally, in~\cref{fig:CutDiffractionEfficiency}(c) we consider the initial momentum width $\Delta \wp = 0.1 \hbar K$ and vary the pulse duration. Raman and Bragg behave again similarly, but for small $\Delta \tau$ off-resonant higher orders start to play a role for double diffraction and change the behavior.

\begin{figure}
	\hspace*{-0.4cm}
	\includegraphics[width =\columnwidth]{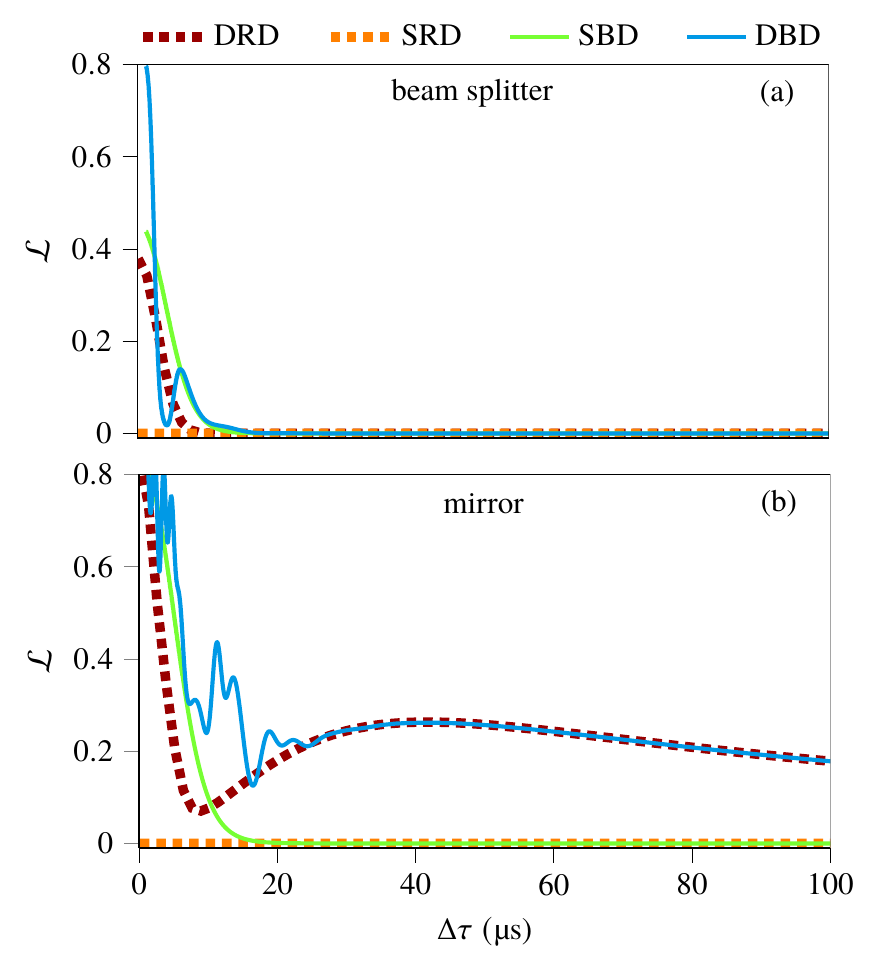}
	\caption{Diffraction losses $\mathcal L$ defined by \cref{eq:defLoss} as a function of the pulse duration $\Delta \tau$ displayed for fixed momentum width $\Delta \wp = 0.1 \hbar K$ and for the four diffraction processes [single Raman: SRD (light orange dashed), single Bragg: SBD (light green solid), double Raman: DRD (dark red dashed), and double Bragg: DBD (dark blue solid)] discussed in this article. For beam splitters (a), losses are only relevant for $\Delta \tau < \SI{20}{\micro \second}$ except for single Raman, which experiences no losses at all. For mirror pulses (b), diffraction losses appear in the case of single Bragg diffraction for small $\Delta \tau$. However, in double diffraction we find additional losses for $\Delta \tau > \SI{20}{\micro \second}$.}
	\label{fig:DiffLosses}	
\end{figure}

\section{\label{sec:DiffractionLosses} Diffraction losses}
\begin{figure*}
	\includegraphics[width =\linewidth]{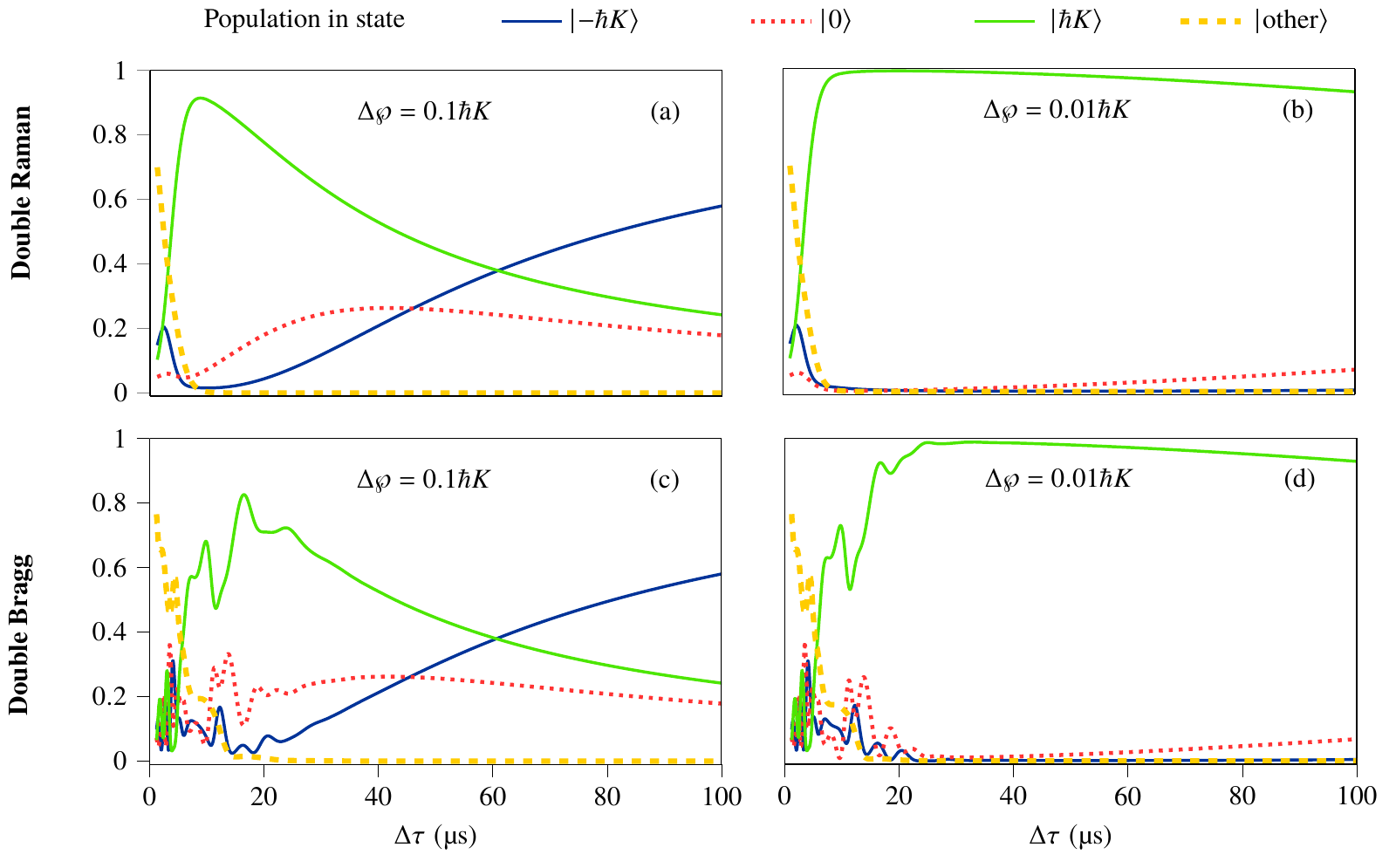}
	\caption{Population in states $\ket{-\hbar K}$, $\ket{0}$, $\ket{\hbar K}$ and off-resonant higher orders $\ket{\mathrm{other}}$ after a mirror pulse for double Raman (top) and double Bragg (bottom). For a broad momentum distribution $\Delta \wp = 0.1 \hbar K$ shown in (a) and (c) a significant part of the population is lost to the intermediate state $\ket{0}$ and off-resonant higher orders are only populated for short pulse durations. Whereas the latter is also true for narrow momentum distributions with $\Delta \wp = 0.01 \hbar K$ illustrated in (b) and (d), we see for intermediate times an almost perfect transfer of the population and a suppression of the losses to the intermediate state.}
	\label{fig:EfficiencyLosses}
\end{figure*}

\begin{figure}
	\includegraphics[width = 220.4888pt]{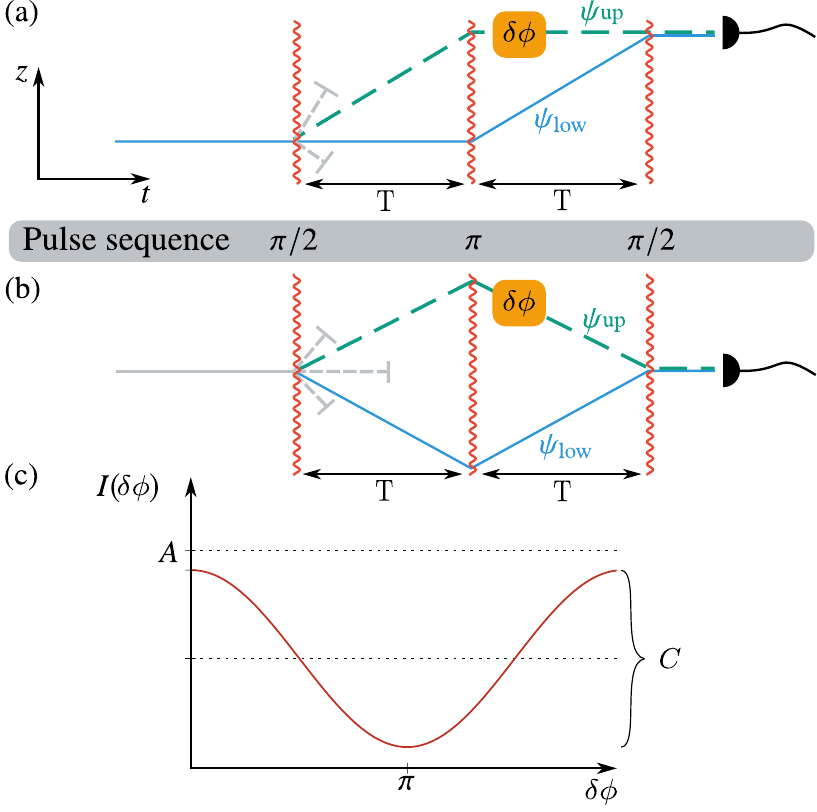}
	\caption{Resonant (solid blue and dashed green) and spurious (dashed gray) paths in a Mach-Zehnder atom interferometer configuration for single (a) and double (b) diffraction. Through a phase scan, we are able to obtain an interference signal $I(\delta \phi)$ in panel (c) with amplitude $A$ and contrast $C$.}
	\label{fig:InterferometerSetup}
\end{figure}

In this section, we investigate why mirrors in double diffraction have a lower efficiency than in single diffraction. We consider as the initial state a Gaussian momentum distribution with a width $\Delta \wp=0.1\hbar K$ and calculate the final population which is neither in the initial nor in the target state. In the following, we refer to this quantity as ``losses'' $\mathcal L$ and define it as
\begin{equation}
\mathcal L \equiv 1 - \int\limits_\mathcal{I} \mathrm{d} p_\mathrm{f} ~\Big| \psi_{\mathrm f}(p_\mathrm{f})\Big|^2,
\label{eq:defLoss}
\end{equation}
where the choice of the integration interval $\mathcal{I}$ depends on the process under consideration. 

We define a beam splitter as a pulse which creates an equal superposition of two momentum states. More specifically, a single-diffraction beam splitter is given by $\ket{0} \rightarrow (\ket{0}+\ket{\hbar K})/\sqrt{2}$ and we choose the integration interval $\mathcal{I}=[-\hbar K/2,3\hbar K/2]$. However, a double-diffraction beam splitter is given by $\ket{0} \rightarrow (\ket{-\hbar K}+\ket{\hbar K})/\sqrt{2}$ with an integration interval $\mathcal{I}=[-3\hbar K/2,3\hbar K/2]$. Hence, $\mathcal{L}$ describes losses to off-resonant higher orders and not losses that arises from a small resonance width and velocity selectivity. 

Losses for beam-splitter processes are shown in \cref{fig:DiffLosses}(a). As expected, in single Raman diffraction they are negligible. We find for single Bragg as well as double Raman again higher-order losses for small pulse durations $\Delta \tau$ and in double Bragg diffraction additionally off-resonant couplings. This behavior changes drastically when we consider mirror pulses in the following.

We recall that a single-diffraction mirror pulse is given by the resonant transition $\ket{0} \rightarrow \ket{\hbar K}$ with the integration interval $\mathcal{I}=[-\hbar K/2,3\hbar K/2]$ and a double mirror by the transition $\ket{-\hbar K} \rightarrow \ket{\hbar K}$ with the integration interval $\mathcal{I}=[-3\hbar K/2,-\hbar K/2] \cup [\hbar K/2,3\hbar K/2]$, i.e. around the initial and final state. Figure~\ref{fig:DiffLosses}(b) shows the losses for mirror pulses. Off-resonant higher-order losses appear again for small $\Delta \tau$ except for single Raman. For larger $\Delta \tau$, the losses decrease for single Bragg, but this is no longer the case for double diffraction. These additional losses substantially reduce the efficiency of double-diffraction mirrors for broad momentum distributions. Off-resonant higher orders can only be populated for small $\Delta \tau$ and thus, they have to be caused by another effect, which we discuss next. 

A double-diffraction mirror can be regarded as a sequence of two resonant processes $\ket{-\hbar K} \rightarrow \ket{0} \rightarrow \ket{\hbar K}$. However, as one considers initial momenta slightly away from resonance, the amplitude of the target state for a mirror pulse starts to decrease. This fact is mainly due to an increasing amplitude of diffraction to the intermediate state close to $\ket{ 0}$ rather than an increasing amplitude of undiffracted atoms in the initial state, as it would be the case for single diffraction. This is a feature, in presence of a small detuning, of the generalized Rabi oscillations for a three-state system that arise in double diffraction. 

In \cref{fig:EfficiencyLosses} we compare the population of the states $\ket{-\hbar K}$, $\ket{0}$, $\ket{\hbar K}$ and off-resonant higher orders for different widths of the input state. Raman is presented at the top, Bragg at the bottom. For a width $\Delta \wp = 0.1\hbar K$ shown in (a) and (c) we observe that only for short pulse durations higher-order momentum states are populated. For longer pulse durations the decrease of efficiency is mainly caused by losses to the intermediate state $\ket{0}$.
	
For sufficiently narrow momentum distributions such as $\Delta \wp = 0.01\hbar K$ illustrated in (b) and (d) these losses to the intermediate state can be significantly suppressed because the dynamics reduces to a resonant three-level system. For momentum eigenstates it was already shown that the population of the intermediate state can be avoided \cite{Giese}. Whereas narrow distributions such as Bose-Einstein condensates can be diffracted with a high efficiency, broader distributions like thermal atoms suffer from significant losses.

Large-momentum transfer techniques of sequential pulses can be conveniently combined with double diffraction, where after an initial beam-splitter pulse, sequential $\pi$ pulses are used to increase the momentum splitting even further \cite{PhysRevLett.114.063002}. For such sequential $\pi$ pulses, losses are less important because they act like single diffraction, see \cref{fig:DiffLosses}(b), on each arm separately due to a Doppler detuning. However, the efficiency of the corresponding composite mirror pulse is still limited by the central double-diffraction mirror, highlighting the need for alternative diffraction schemes for broad momentum distributions. 

\begin{figure*}
	\includegraphics[width =.93\linewidth]{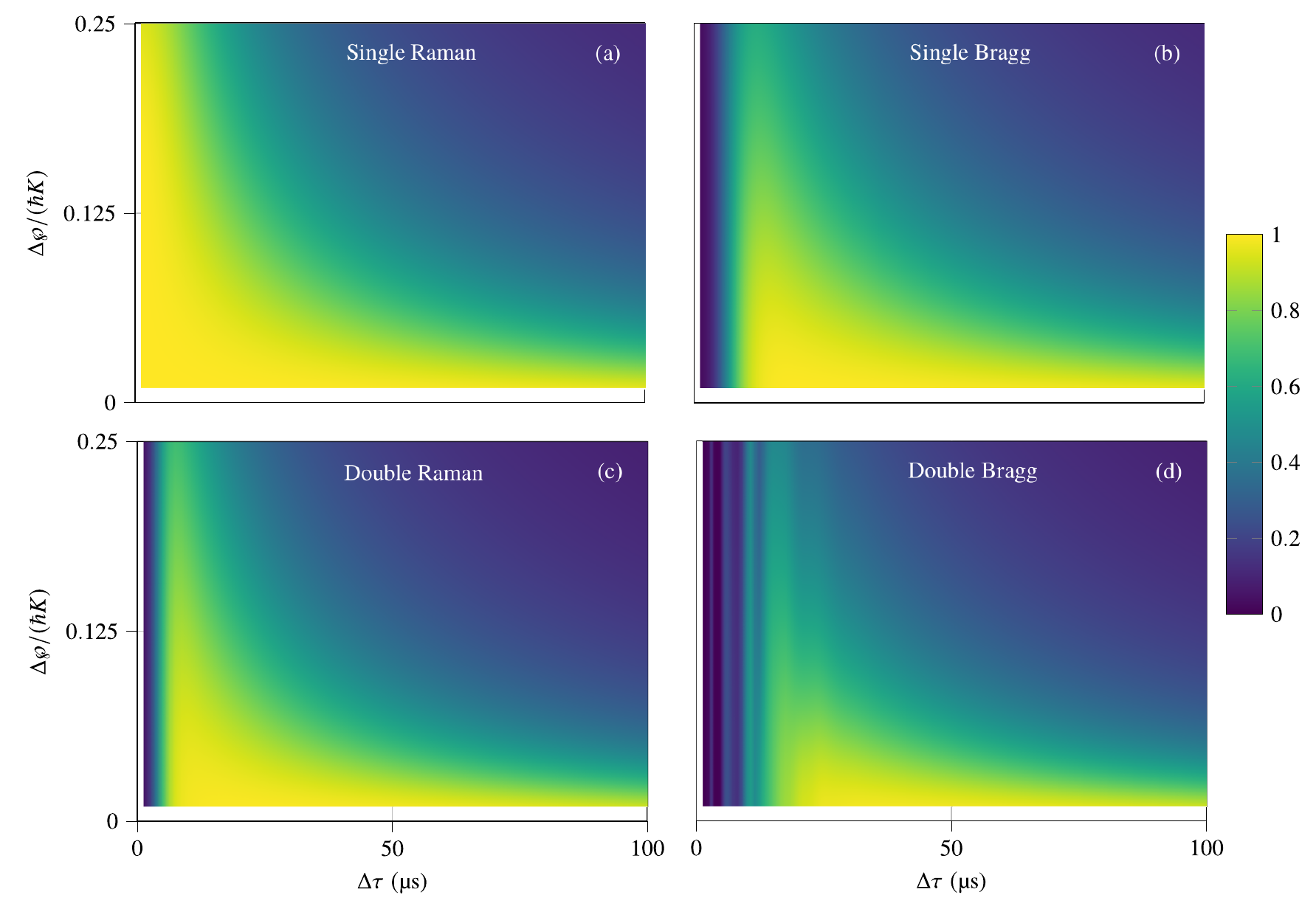}
 	\vspace*{-0.1cm}
	\caption{Amplitude $A$ of a simple Mach-Zehnder interferometer as a function of the momentum width $\Delta \wp$ of the initial wave packet and the pulse duration $\Delta \tau$. We observe for single Raman (a), single Bragg (b), double Raman (c) and double Bragg (d) how the imperfections of the diffracting pulses influence the interference signal. In particular, the effect of the Bragg regime as well as quasi-resonances are reflected in the amplitude of the signal.}
	\label{fig:amplitude}
\end{figure*}

\begin{figure}
	\includegraphics[width =.93\linewidth]{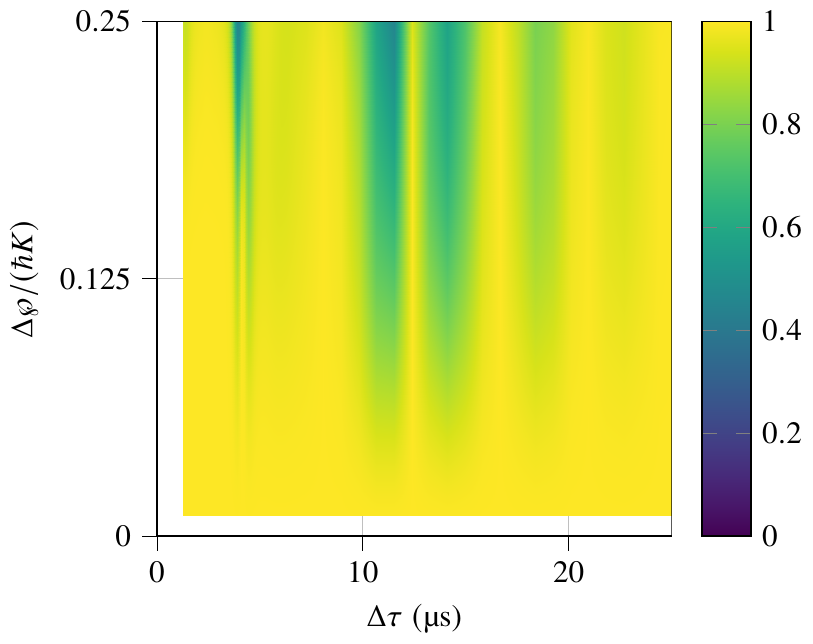}
	 \vspace*{-0.1cm}
	\caption{Interferometric contrast $C$ of a double-Bragg Mach-Zehnder interferometer as a function of the momentum width $\Delta \wp$ of the initial wave packet and the pulse duration $\Delta \tau$. The contrast is close to unity except for parameters where quasi-resonances dominate the behavior, where it drops to 0.5. However, because of the impracticality of blow-away pulses for Bragg, additional contributions from spurious paths, which are not included in this treatment, may dominate the contrast.}
	\label{fig:contrast}
\end{figure}

\begin{figure*}
	\includegraphics[width =\linewidth]{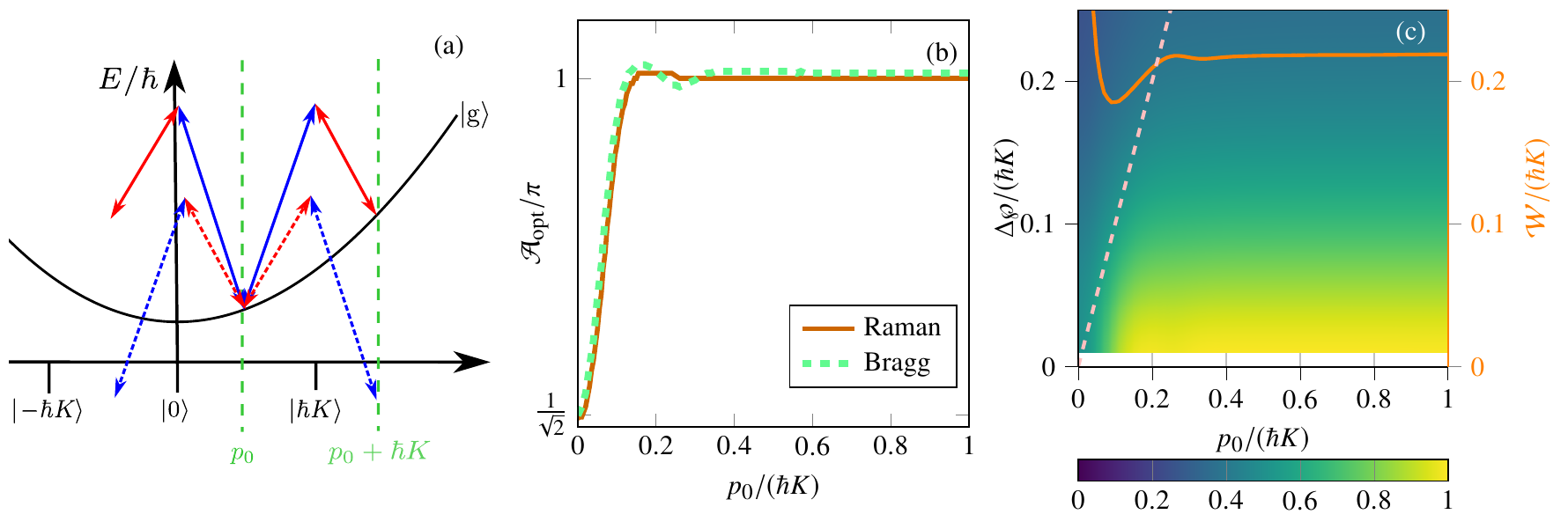}
	\caption{Energy-momentum diagram for double Bragg diffraction (a) for an initial momentum offset $p_0$ resonantly connected to $p_0 + \hbar K$. Pulse area $\mathcal{A}_\text{opt}$ defined for single diffraction (b) leading to an optimal diffraction efficiency for an initial \emph{momentum eigenstate} $\Ket{p_0}$ and a pulse duration $\Delta \tau = 37.5\,$\textmu s, comparing Raman and Bragg diffraction. We compute the efficiency for different initial momenta and Gaussian widths $\Delta \wp$ and show in (c) the results for Bragg diffraction with a pulse duration of $\Delta \tau = \SI{37.5}{\micro \second}$ using the optimal pulse area from (b). For $p_0 = 0$ and small $\Delta \wp$ we observe an efficiency of $0.5$ corresponding to a double-diffraction beam-splitter process. For increasing $p_0$, the efficiency approaches unity and thus a single-diffraction mirror process emerges. We also display the resonance width $\mathcal{W}$ of the diffraction process as a function of $p_0$ in this density plot (orange line) and observe that it is one limiting factor of the efficiency. The straight line $p_0=\Delta \wp$ (see dashed pink line) in the plot demonstrates that a significant fraction of the atoms undergo double instead of single diffraction whenever the momentum distribution has significant overlap with the degenerate momentum ($p_0=0$), i.e. $\Delta \wp > p_0$, as long as the resonance is wide enough. The corresponding behavior for Raman diffraction is similar. Effects on phases, caused for example by two-photon light shifts, cannot be observed in the efficiency.}
	\label{fig:SingleToDouble}	
\end{figure*}

\section{\label{sec:Interferometer} Interferometric contrast}
Since the effect of losses and an imperfect efficiency potentiate for a sequence of pulses, an interferometer is even more affected by the processes described above. Additionally, imperfect beam splitters and mirrors lead to more than two paths. Some of the spurious paths might end up in the detected exit port, leading to a loss of contrast through a beating of multiple interference signals or a background signal. In our discussion, we focus on a Mach-Zehnder interferometer consisting of a sequence of beam-splitter, mirror, and beam-splitter pulses separated by a time $T$ as shown for both single and double diffraction in \figsref{fig:InterferometerSetup} (a) and (b).

Whereas in a realistic interferometer additional spurious paths (dashed) might end in the considered exit port in our present discussion, we focus in our discussion only on the two main paths (solid) and disregard the spurious ones (indicated by blocked paths in the figures). Hence, the signal will not be affected by the beating of multiple interference signals or a background signal, even though such effects can in principle be obtained from our results as well. Instead, here we only investigate the effect of the loss of efficiency.

	In double Raman diffraction some of the spurious paths can be removed by means of suitable blow-away pulses \cite{peters2001high}, but in general spurious paths can play a relevant role in certain regimes of Bragg and Raman diffraction. In fact, building up on methods introduced in Ref.~\cite{Roura_2014}, it is possible to develop a code that fully takes into account the contributions of spurious paths and can be applied to arbitrarily long interferometer times. This will be presented in some future publication, whereas here we will only consider the two main paths depicted in \cref{fig:InterferometerSetup}.

	Given an initial momentum eigenstate, the phase accumulated due to free evolution between two laser pulses in one arm is $\mathrm{exp} \left( - \mathrm i (p^2 / (2m\, \hbar))\, T \right)$, where $p$ is the corresponding momentum in that segment. In addition, in interferometers involving Raman pulses, and hence changes of the internal state, an extra phase $\mathrm{exp} \left(- \mathrm i \omega_\text{eg} T \right)$ is accumulated in those segments where the atoms are in the excited state.
	Disregarding possible contributions from spurious paths simplifies significantly the computation of the state evolution in the atom interferometer.
	Indeed, in the Mach-Zehnder interferometer with double-diffraction pulses of \cref{fig:InterferometerSetup}(b), the internal state is at any time the same for the two arms and $p^2 = (-p)^2$ as well. Therefore, the contributions of the associated phases to the phase shift $\delta\phi$ between the two arms of the interferometer cancel out. Similarly, in the Mach-Zehnder interferometer with single-diffraction pulses of  \cref{fig:InterferometerSetup}(a), the phase accumulated in one arm between the first and second pulses is identical to the phase accumulated between the second and third pulses in the other arm and vice versa, so that these phase-shift contributions also cancel out. Moreover, the same cancelations hold for an arbitrary initial wave packet, which can always be regarded as a linear superposition of momentum eigenstates.

	Thus, to obtain the interference signal at each exit port, it is sufficient to multiply specific elements of the transition functions $G_{\Delta \tau}^{\text{(M/BS)}}$, as shown in \cref{app:Wavefunctions}, when considering the evolution of the initial wave packet $\psi_{\mathrm{i}}(p)$ along each interferometer arm. In addition, we include an extra phase $\delta\phi$ in one of the two arms which can account, for instance, for the effects of a uniform gravitational field. Indeed, when transforming to a freely falling frame, the dynamics of the atomic wave packets reduces to that depicted in \cref{fig:InterferometerSetup} and one simply gets an additional phase-shift contribution $\delta\phi = k_\text{eff}  g\, T^2$ from the transformation of the laser phases to the freely falling frame \cite{roura2018gravitational} plus terms of order $(\Delta \tau / T)$ due to the finite pulse duration \cite{antoine_matter_2006,bertoldi_phase_2019,cheinet_measurement_2008}. Here $g$ corresponds to the projection of the gravitational acceleration onto the beam direction in the laboratory frame and the effective momentum transfer $\hbar k_\text{eff}$ is given by $k_\text{eff} = K$ for single and $k_\text{eff} = 2\,K$ for double diffraction.

	Even though the aforementioned phase caused by linear accelerations arises for both single and double diffraction, it does not account for the detuning associated with the shift of the momentum distribution experienced in the laboratory frame between the pulses. For single diffraction along the vertical direction the gravitational acceleration quickly leads to a substantial deviation from the resonance condition given by \cref{eq:ResCondRaman} or \cref{eq:ResCondBragg} and it is necessary to chirp the frequency difference $\Delta \omega$ at a certain rate $\alpha$ so that it stays close to resonance. The phase shift becomes then $\delta \phi = k_\text{eff} (g-\alpha) T^2$, and one can scan the phase shift $\delta \phi$ by slightly varying the chirping rate around the resonance value $\alpha = g$ \cite{peters2001high}.

	However, for double diffraction in a retroreflective geometry such a chirping is impossible due to the symmetry of the setup. However, this problem can be circumvented by using three laser frequencies as experimentally demonstrated for Raman~\cite{malossi_double_2010}. Alternatively, for nearly horizontal beams the projection of the gravitational acceleration onto the beam direction can be small enough so that it results into almost no deviation from resonance. Slightly changing the angle between the beam and the horizontal direction can then be employed to scan the phase shift $\delta \phi$ as done in Ref.~\cite{ahlers_double_2016}.

By multiplying the respective elements of the transition function, as indicated in \cref{app:Wavefunctions}, we obtain the wave functions $\psi_{\mathrm{up}}$ and $\psi_{\mathrm{low}}$ that have been propagated along the upper and lower arm, respectively. The interference signal then takes the form
\begin{equation}
\begin{split}
I(\delta\phi) &= \int \limits_{-\hbar K/2}^{\hbar K/2} \mathrm{d}p_\mathrm{f} \, \left| \psi_\mathrm{up}(p_\mathrm{f}) \, \mathrm{e}^{\mathrm{i} \delta\phi} +\psi_\mathrm{low}(p_\mathrm{f}) \right|^2 \\
&= \frac{A}{2} \left(1 + C \cos \delta\phi\right)
\end{split}
\end{equation}
where $A$ and $C$ denote the amplitude and contrast of the signal,
and the expressions for $\psi_\mathrm{up}(p_\mathrm{f})$ and $\psi_\mathrm{low}(p_\mathrm{f})$ are given by \cref{eq:psi_SD,eq:psi_DD} for the two diffraction mechanisms. The amplitude $A$ characterizes losses to spurious paths, but more in general it can also be influenced by contributions of spurious paths to the same exit port. However, the contrast $0 \leq C \leq 1$ may decrease due to imbalances in the diffraction efficiencies for the two arms and to unequal distortions of the atomic wave packets caused by the laser pulses as they evolve along both arms. Full contrast ($C=1$) is recovered when these effects are negligible.

In our simulations, we scan the phase shift $\delta\phi$ from $0$ to $2\pi$ for different pulse durations and momentum widths of the initial distribution. A typical interference signal is shown in Fig.~\ref{fig:InterferometerSetup}(c). From these signals, we obtain the amplitude through $A= \operatorname{max}[I(\delta\phi)] + \operatorname{min}[I(\delta\phi)]$ and present the results in the density plots of Fig.~\ref{fig:amplitude} for both single and double as well as Raman and Bragg diffraction. In these plots we can directly infer the influence of imperfect beam splitters and mirrors on the interference signal and observe effects similar to those obtained for the individual diffracting elements: single Raman has the largest amplitude over the widest range of parameters, double Raman and single Bragg have to performed in a Bragg-type regime, and for double Bragg quasi-resonances lead to an increase of amplitude for specific parameters. Since double-diffraction mirrors introduce a significant loss of population to the intermediate state for broad momentum distributions, they also limit the overall signal. However, the amplitude of single diffraction is not limited by such effects and is therefore higher even for broad momentum distributions.

In our framework the contrast is given by \mbox{$C= \left( \operatorname{max}[I(\delta\phi)] - \operatorname{min}[I(\delta\phi)] \right) / A$} and is reduced by asymmetric wave packed distortions on the two main paths caused by imperfect diffraction. Since such distortions are small for all cases except for double Bragg, no loss of contrast can be observed throughout the range of investigated parameters. Although the quasi-resonances increase the efficiency of double Bragg diffraction, they also lead to an asymmetry between the wave packets and by that to a significant loss of contrast for the double Bragg case. We show the contrast as a function of pulse duration and momentum width in Fig.~\ref{fig:contrast} and see that it drops significantly in the parameter regime where quasi-resonances are dominant. In fact, we observe a drop in contrast to about 0.5 for broad distributions. Note that we have restricted in this plot the range of pulse durations to the quasi-Bragg regime.

However, especially for double Bragg diffraction the interference signal will be dominated by other spurious effects that are neglected in our treatment. Because no blow-away pulses can be applied to Bragg (and single Raman), atoms that remain unaffected by all three pulses will lead to a third path that ends in the exit port under consideration. If this wave packet has no overlap in momentum space with the diffracted components of the main paths (for example the atoms that are unaffected due to velocity selectivity), it will lead to a background population that alters the value of $A$ and significantly reduces the contrast. However, if this contribution has an overlap in momentum with the main paths, it leads to a beating of three interferograms and the signal will be more complex so that it cannot be characterized by a simple amplitude and contrast anymore. Since we have neglected spurious paths in the present simulation, such effects, which can actually dominate the signal for double Bragg in certain parameter regimes, are not included in our analysis, but will be discussed in future work.

\section{\label{sec:Transition} From double to single diffraction}
As discussed in \secref{sec:Introduction}, single diffraction is conventionally performed in a retroreflective setup where one Doppler-detuned grating can be neglected and the process can be modeled by the interaction with a single grating. In this section, we demonstrate the transition from double to single diffraction in a retroreflective setup as the Doppler detuning increases for one grating while keeping the other one resonant. 

In fact, for a momentum $p_0 \neq 0$ one of the two transitions becomes Doppler detuned. For small detunings, double diffraction is asymmetric and for large detunings, one of the two transitions is completely suppressed. Even though in this case the off-resonant transitions lead to no diffraction, and the efficiency resembles the one of single diffraction, they can nevertheless have an effect on the phase of the atom, through two-photon light shifts~\cite{gauguet_off-resonant_2008,carraz_phase_2012,gillot_limits_2016,giese_light_2016}.

We show the transition from double to single diffraction in a retroreflective geometry by increasing the initial momentum $p_0$ from $0$ to $\hbar K$. For each momentum, we adjust the resonance condition to
\begin{equation}
    \Delta \omega = \omega_{\mathrm{eg}} + \omega_K + p_0 K / m
\end{equation}
for Raman and
\begin{equation}
\Delta \omega = \omega_K + p_0 K / m
\end{equation}
for Bragg. 

We see in~\cref{fig:SingleToDouble}(a) for the case of Bragg that the additional Doppler detuning in the resonance condition is necessary to resonantly connect the momentum states $\ket{p_0}$ and $\ket{p_0 + \hbar K}$. We use these resonance conditions and the differential equations for double diffraction given in \cref{App:DGLs}. 

Since the Rabi frequencies of single and double diffraction differ by a factor of $\sqrt{2}$, we use the definition of the pulse area for single diffraction to obtain numerically the area $\mathcal{A}_\text{opt}$ for which the resonant \emph{momentum eigenstate} $\ket{p_0}$ is diffracted most efficiently to $\ket{p_0 + \hbar K}$. These pulse areas for Raman and Bragg are presented in Fig. 11(b) for a pulse duration of $\Delta \tau = 37.5\,$\textmu s. As expected, for no initial momentum the optimal area is $\pi / \sqrt{2}$, which corresponds to a double-diffraction beam splitter. For $p_0 / \hbar K \geq 0.1$ the optimal area reaches $\pi$ which corresponds to a single-diffraction mirror process. Even though Bragg diffraction has a richer structure than Raman diffraction due to simultaneous resonant and off-resonant transitions, the optimal pulse areas behave very similar\footnote{Note that small (numerical) fluctuations of the efficiency lead to slightly different optimal pulse areas.}. 

With these optimal pulse areas $\mathcal{A}_{\mathrm{opt}}(p_0)$ we calculate for each $p_0$ the diffraction efficiency of initial distributions of different widths $\Delta \wp$ centered around $p_0$ where the integration interval is modified accordingly. The only noteworthy difference in efficiency between Raman and Bragg appears in the region $p_0/\hbar K<0.2$ and small $\Delta \wp$ and is below $5 \%$. Therefore in~\cref{fig:SingleToDouble}(c), we show  only the Bragg case. We observe that for a sufficiently narrow distribution the efficiency approaches unity already for small Doppler detunings $p_0/\hbar K \gtrsim 0.1$, even though the explicit transition from double to single depends on the pulse duration $\Delta \tau$. As expected the efficiency decreases for increasing $\Delta \wp$. 

To get an intuitive understanding of the dominant effect, and in analogy to \cref{sec:DiffractionLosses}, we calculate the resonance width $\mathcal{W}$ of the diffraction process through the FWHM of the transition function as a function of $p_0$. We present the result in the density plot of \cref{fig:SingleToDouble}(c) (see orange line). If the momentum distribution is wider than the resonance width, then the efficiency decreases rapidly. However, for small $p_0$ the efficiency seems to be bounded by another quantity: The straight line $p_0=\Delta \wp$ (dashed pink line) demonstrates that if the momentum distribution has overlap with the degenerate momentum ($p_0=0$), i.e. $\Delta \wp > p_0$, and the resonance is sufficiently broad, then a significant fraction of the atoms undergo double instead of single diffraction, leading to a decrease of efficiency.

\Cref{fig:SingleToDouble}(b) also reveals that the optimal pulse area for the Doppler-detuned atomic sample is slightly larger than $\pi$. This effect shows that for a pulse duration of $\Delta \tau = \SI{37.5}{\micro \second}$ spurious diffraction orders change the effective Rabi frequency \cite{Giese}.

\section{\label{sec:Conclusion} Conclusions}
Our article provides a detailed study of single and double diffraction for both mechanisms, Raman and Bragg. In particular, we have shown that single and double diffraction can both be realized in a retroreflective setup, and that already a Doppler detuning corresponding to a momentum $\gtrsim 0.1\,\hbar K$ is sufficient to suppress the effect of a second grating, turning double into single diffraction. Alternatively, chirping of the laser frequencies together with finite speed-of-light effects can also lift the degeneracy of the double diffraction process and lead to a preferred direction of diffraction \cite{Perrin_chriping_2019}.

Moreover, we have compared the diffraction efficiencies of both mechanisms for the same parameters and observed that for single Bragg as well as for double diffraction in general it is not possible to realize a Bragg-type regime with short pulse durations. In this sense, single Raman diffraction has the unique property that it can be performed also for intense pulses without a significant loss of efficiency. However, beyond these short pulse durations we have observed no significant difference between Raman and Bragg diffraction for a wide range of the parameter regime.

These insights have consequences for the velocity selectivity of the pulses: in principle, the resonance width of the process increases for shorter times, until one leaves the Bragg-type regime (except for single Raman). Hence, for each momentum width there exists a unique optimal pulse duration with a pulse area of $\pi$. Double Bragg constitutes an exception, as quasi-resonances arising for particular pulse durations allow to increase the diffraction efficiency of broad momentum distributions significantly.

Moreover, for broad momentum distributions we have demonstrated that all double-diffraction mirrors are less efficient than their single-diffraction counterparts irrespectively of the diffraction mechanism, because there is a considerable loss of atoms into the intermediate momentum state, a feature that does not exist in single diffraction.

Throughout our article we have focused on the number of diffracted atoms as a measure of the quality of the diffraction process. However, for increasingly sensitive ground experiments as well as future space missions with atom-interferometric capabilities the influence of diffraction regimes, geometries and mechanisms on the phases are crucial. For this reason these topics are part of our future research program. 

\begin{acknowledgements}
We thank C.~M. Carmesin, A. Friedrich, M. Gebbe and C. Schubert for fruitful discussions.
This project was generously supported by the German Aerospace Center (Deutsches Zentrum für Luft- und Raumfahrt, DLR)
 with funds provided by the Federal Ministry for Economic
Affairs and Energy (Bundesministerium für Wirtschaft und Energie, BMWi) under the Grant Nos. 50WM1556 (QUANTUS
IV), 50WM1956, 50WM1952 (QUANTUS V), 50WP1705, 50WP1700 (BECCAL) and 50RK1957 (QGYRO).
The research of the IQ$^\mathrm{ST}$ is financially supported by the
Ministry of Science, Research and Arts Baden-W\"{u}rttemberg (Ministerium für Wissenschaft, Forschung und Kunst Baden-Württemberg). The research of the Institut für Quantenoptik financially supported by the CRC 1227 DQmat within the projects A05 and B07, the EXC 2123 Quantum Frontiers within the research units B02 and B05, the QUEST-LFS, the Association of German Engineers ()Verein Deutscher Ingenieure, VDI) with funds provided by the Federal Ministry of Education and Research (Bundesministerium für Bildung und Forschung, BMBF) under Grant No. VDI 13N14838 (TAIOL), and ``Nieders{\"a}chsisches Vorab'' through the ``Quantum- and Nano-Metrology (QUANOMET)'' initiative within the project QT3 as well as through ``F{\"o}rderung von Wissenschaft und Technik in Forschung und Lehre'' for the initial funding of research in the new DLR-SI Institute. W.~P.~S. is most grateful to Texas A\&M University for a Faculty
Fellowship at the Hagler Institute for Advanced Study at the Texas
A\&M University as well as to the Texas A\&M AgriLife Research for
its support. 
S.~H. and J.~J. contributed equally to this work.

\end{acknowledgements}

\appendix

\section{Full equations}
\label{App:DGLs}
In this Appendix we present the differential equations describing the four diffraction processes discussed in our article. Here, we do not focus on a particular resonance condition but consider the general case. Indeed, by choosing an appropriate frequency difference $\Delta \omega$ between the lasers, one can implement other resonant transitions. 

The derivation of these equations uses the rotating wave approximation~\cite{schleich_quantum_2001} and the adiabatic elimination of the optically excited state~\cite{bernhardt_coherent_1981,marte_multiphoton_1992,brion_adiabatic_2007}. The dynamics is in an interaction picture with respect to the free evolution of the atoms, and we assume vanishing laser phases.

The equations~\cite{moler}
\begin{subequations}
\begin{equation}
\begin{split}
\mathrm i \dot g_n & =-\Omega \, \mathrm e^{-\mathrm i \omega_\mathrm{D}  t} \, \mathrm e^{-\mathrm i [\omega_\mathrm{eg}-\Delta \omega +\omega_{\mathrm{AC}}+(1+2n)\omega_K]t} \,  e_{n+1}
\end{split}
\end{equation}
and
\begin{equation}
\begin{split}
\mathrm i \dot e_{n+1} & =-\Omega \, \mathrm e^{\mathrm i \omega_\mathrm{D}  t} \, \mathrm e^{-\mathrm i [-\omega_{\mathrm{eg}}+\Delta \omega-\omega_{\mathrm{AC}}-(1+2n)\omega_K]t} \, g_n,
\end{split}
\end{equation}
\label{eq:FullSingleRaman}
\end{subequations}
for single Raman diffraction with the probability amplitudes $g_n \equiv g(p + n \hbar K)$ and $e_n \equiv e(p + n \hbar K)$ of the ground state and excited state in momentum representation form a closed system of coupled differential equations. 

The intensity and the pulse shape of the grating determine the coupling strength $\Omega = \Omega(t)$. The frequency difference of the two internal states is given by $\omega_{\mathrm{eg}}$. The AC Stark shift is denoted by $\omega_{\mathrm{AC}}$ and the recoil frequency by $\omega_{K}$. The Doppler frequency $\omega_\mathrm{D}$ acts as a detuning. 

An appropriate choice of $\Delta \omega$, as given by \cref{eq:ResCondRaman}, allows us to neglect the second exponent in \cref{eq:FullSingleRaman} for resonant transitions while off-resonant transitions oscillate with multiples of $\omega_K$. In this way the dependence on $\omega_{\mathrm{eg}}$ drops out of the exponent. Note, however, that terms with frequencies involving higher multiples of $\omega_{\mathrm{eg}}$ have been neglected in \cref{eq:SingleRaman,eq:FullSingleRaman} as a result of the rotating wave approximation. This is justified in our case because we typically have $\omega_{\mathrm{eg}} \sim 2 \pi \times 7$~GHz and even for the shortest pulses considered here, with $\Delta \tau \sim 1$\textmu s, the condition $\omega_{\mathrm{eg}} \Delta \tau \gg 1 $ is amply fulfilled. 
\begin{widetext}
Similarly, single Bragg diffraction is described by the recurrence relation~\cite{Giese}
\begin{equation}
\mathrm i \dot g_n = - \Omega \, \mathrm e^{- \mathrm i \omega_\mathrm{D}  t} \, \mathrm e^{\mathrm i [\Delta \omega - (2n + 1) \omega_K] t} \, g_{n + 1}
- \Omega \, \mathrm e^{\mathrm i \omega_\mathrm{D}  t} \, \mathrm e^{- \mathrm i [\Delta \omega - (2n - 1) \omega_K] t} \, g_{n - 1},
\end{equation}
which in contrast to single Raman diffraction, given by~\cref{eq:FullSingleRaman}, is not closed. 

The double Raman-equations contain the terms of~\cref{eq:FullSingleRaman} and contributions corresponding to the additional laser pairs for perfect orthogonal polarizations. The resulting coupled differential equations
\begin{subequations}
\begin{equation}
 \mathrm i  \dot g_n =  - \Omega \, \mathrm e^{-\mathrm i \omega_\mathrm{D}  t} 
                      \, e^{-\mathrm i [\omega_{\mathrm{eg}}-\Delta \omega +\omega_{\mathrm{AC}}+(1+2n)\omega_K]t} \, e_{n+1}
                      - \Omega \, \mathrm e^{\mathrm i \omega_\mathrm{D}  t} 
                      \, e^{-\mathrm i [\omega_{\mathrm{eg}}-\Delta \omega +\omega_{\mathrm{AC}}+(1-2n)\omega_K]t} \, e_{n-1}
\end{equation}
and
\begin{equation}
\mathrm i \dot e_{n+1} = - \Omega \, \mathrm e^{-\mathrm i\omega_\mathrm{D}  t} 
                          \, e^{-\mathrm i [-\omega_{\mathrm{eg}}+\Delta \omega-\omega_{\mathrm{AC}}+(3+2n)\omega_K]t} \, g_{n+2}
                        - \Omega \, \mathrm e^{\mathrm i\omega_\mathrm{D}  t}
                          \,e^{-\mathrm i [-\omega_{\mathrm{eg}}+\Delta \omega-\omega_{\mathrm{AC}}-(1+2n)\omega_K]t} \, g_{n}
\end{equation}
\end{subequations}
show that simultaneous diffraction in both directions is now possible. A truncated version of these equations can be found in Ref.~\cite{ThesisLeveque}. 

Simultaneous diffraction in both directions also occurs for double Bragg diffraction which is described by the equations~\cite{Giese}
\begin{equation}
 \mathrm i \dot g_n = - \Omega \, \mathrm e^{- \mathrm i \omega_\mathrm{D}  t} 
                       \left[\mathrm e^{- \mathrm i [\Delta \omega + (2n + 1) \omega_K] t} + \mathrm e^{\mathrm i [\Delta \omega - (2n + 1) \omega_K] t}\right] \, g_{n + 1} 
                      - \Omega \, \mathrm e^{\mathrm i \omega_\mathrm{D}  t}  
                       \left[\mathrm e^{\mathrm i [\Delta \omega + (2n - 1) \omega_K] t} + \mathrm e^{- \mathrm i [\Delta \omega - (2n - 1) \omega_K] t}\right] \, g_{n - 1}.
\end{equation}
Here, states can simultaneously couple both resonantly and off-resonantly.

\section{Wave packet components in an interferometer exit port}
\label{app:Wavefunctions}
We present in this appendix the calculation of the wavepacket components in the exit port of a Mach-Zehnder interferometer. For that, we assume that the initial wave packet $\psi_\text{i}$ is centered around the momentum $p_0=0$. As explained in \secref{sec:Interferometer}, for the interferometers shown in \cref{fig:InterferometerSetup} the phases associated with the free evolution and changes of internal state (for Raman transitions) along the two arms cancel out and will not be included here.
For Raman diffraction the transition between the internal states $\ket{\mathrm g}$ and $\ket{\mathrm e}$ as well as the momentum transfer are described in momentum representation by the operator 
\begin{equation}
\hat G_{\Delta\tau}^{\text{(BS/M)}}(p_\mathrm{f},p_\mathrm{i}) = \sum\limits_{j,l \in \{\mathrm e,\mathrm g\}} G_{\Delta\tau, jl}^{\text{(BS/M)}}(p_\mathrm{f},p_\mathrm{i})\ket{j}\bra{l}.
\end{equation}
 However, for Bragg it reduces to the diagonal elements proportional to $\ket{\mathrm g} \bra{\mathrm g}$ that directly describe the transition function defined in \secref{sec:NumericalTreatment}.
	For a state that is prepared and detected in $\ket{\mathrm g}$, the evolution along each interferometer arm is then obtained by multiplying the respective elements of the transition functions $G_{\Delta\tau, jl}^{\text{(BS/M)}}(p_\mathrm{f},p_\mathrm{i})$. Neglecting spurious paths, in single diffraction we find for the upper arm
\begin{subequations}
	\begin{equation}
	\psi_\mathrm{up}(p_\mathrm{f}) =  \int\limits_{-\hbar K/2}^{\hbar K/2} \mathrm{d}p_2 \int\limits_{-\hbar K/2}^{ \hbar K/2} \mathrm{d}p_1 \int\limits_{\hbar K/2}^{3\hbar K/2} \mathrm{d}p_\mathrm{i}  \,  G_{\Delta \tau,\mathrm{gg}}^{\mathrm{(BS)}}(p_\mathrm{f},p_2) \, G_{\Delta \tau,\mathrm{ge}}^{\mathrm{(M)}}(p_2,p_1)  \, G_{\Delta \tau,\mathrm{eg}}^{\mathrm{(BS)}}(p_1,p_\mathrm{i}) \, \psi_\mathrm{i}(p_\mathrm{i})
	\end{equation}
	where the respective transition elements are chosen by the limits of the integrals. Similarly, the lower arm can be calculated by 
	\begin{equation}
	\psi_\mathrm{low}(p_\mathrm{f}) = \int\limits_{-\hbar K/2}^{\hbar K/2} \mathrm{d}p_2 \int\limits_{\hbar K/2}^{3\hbar K/2} \mathrm{d}p_1 \int\limits_{-\hbar K/2}^{\hbar K/2} \mathrm{d}p_\mathrm{i} \, G_{\Delta \tau,\mathrm{ge}}^{\mathrm{(BS)}}(p_\mathrm{f},p_2) \, G_{\Delta \tau,\mathrm{eg}}^{\mathrm{(M)}}(p_2,p_1)  \, G_{\Delta \tau,\mathrm{gg}}^{\mathrm{(BS)}}(p_1,p_\mathrm{i}) \, \psi_\mathrm{i}(p_\mathrm{i}).
	\end{equation}
		\label{eq:psi_SD}
\end{subequations}	
	For double diffraction, the same line of reasoning leads to
\begin{subequations}	
	\begin{equation}
	\psi_\mathrm{up}(p_\mathrm{f}) =  \int\limits_{-\hbar K/2}^{\hbar K/2} \mathrm{d}p_2 \int\limits_{-3\hbar K/2}^{-\hbar K/2} \mathrm{d}p_1 \int\limits_{\hbar K/2}^{3\hbar K/2} \mathrm{d}p_\mathrm{i} \, G_{\Delta \tau,\mathrm{ge}}^{\mathrm{(BS)}}(p_\mathrm{f},p_2) \, G_{\Delta \tau,\mathrm{ee}}^{\mathrm{(M)}}(p_2,p_1)  \, G_{\Delta \tau,\mathrm{eg}}^{\mathrm{(BS)}}(p_1,p_\mathrm{i}) \, \psi_\mathrm{i}(p_\mathrm{i})
	\label{eq:psi_up_DD}
	\end{equation}
	for the upper arm and to
	\begin{equation}
	\psi_\mathrm{low}(p_\mathrm{f}) = \int\limits_{-\hbar K/2}^{\hbar K/2} \mathrm{d}p_2 \int\limits_{\hbar K/2}^{3\hbar K/2} \mathrm{d}p_1 \int\limits_{-3\hbar K/2}^{-\hbar K/2} \mathrm{d}p_\mathrm{i}  \, G_{\Delta \tau,\mathrm{ge}}^{\mathrm{(BS)}}(p_\mathrm{f},p_2) \, G_{\Delta \tau,\mathrm{ee}}^{\mathrm{(M)}}(p_2,p_1)  \, G_{\Delta \tau,\mathrm{eg}}^{\mathrm{(BS)}}(p_1,p_\mathrm{i}) \, \psi_\mathrm{i}(p_\mathrm{i})
	\end{equation}
	\label{eq:psi_DD}
\end{subequations}
	for the lower one. In Bragg diffraction there is no internal transition and thus $G_{\Delta\tau, jl}^{\text{(BS/M)}}(p_\mathrm{f},p_\mathrm{i}) \rightarrow G_{\Delta\tau, \mathrm{gg}}^{\text{(BS/M)}}(p_\mathrm{f},p_\mathrm{i})$.
\end{widetext}

\bibliography{bibliography.bib}

\end{document}